\documentclass{aa}  

\usepackage{graphicx}
\usepackage{hyperref}
\hypersetup{linkcolor=blue,citecolor=blue,colorlinks=true,pdfborderstyle={}}
\usepackage{natbib}
\usepackage{xcolor}
\usepackage{xspace}
\usepackage{xcolor}
\usepackage{amsmath}
\usepackage{amssymb}
\usepackage{multirow}
\usepackage{graphicx}
\usepackage{booktabs}
\usepackage{adjustbox}
\usepackage{subcaption}
\usepackage{enumerate}
\usepackage{enumitem}
\usepackage{float}
\usepackage{soul}
\usepackage{txfonts}

\newcommand{\linkorcid}[1]{\href{https://orcid.org/#1}{\includegraphics[width=8pt]{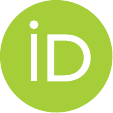}}}
\begin{document}

   \title{Impact of model parameter degeneracy on \\ leptonic radiation models}

   \subtitle{The case of blazar multi-wavelength spectra}

   \author{F. Apel
          \inst{\ref{rub}}
          \and
          A. Omeliukh\inst{\ref{rub}} \linkorcid{0000-0001-8785-9771}
          \and
          A. Franckowiak\inst{\ref{rub},\ref{rapp}}\linkorcid{0000-0002-5605-2219}
          \and
          J. Lederer\inst{\ref{HH}}
          }

    \institute{Ruhr University Bochum, Faculty of Physics and Astronomy, Astronomical Institute (AIRUB), Universitätsstraße 150, 44801 Bochum, Germany, \email{apel@astro.rub.de,omeliukh@astro.rub.de} \thanks{The work was split by the two first authors in equal parts.} \label{rub}
        \and 
             Ruhr Astroparticle and Plasma Physics Center (RAPP Center) \label{rapp}
        \and
            University of Hamburg, MIN Department, Bundesstraße 55, 20146 Hamburg, Germany \label{HH}
            }

   \date{Received XXX; accepted XXX}

 \newcommand{\jl}[1]{\textcolor{red}{#1}}
  \abstract
   {Leptonic one-zone radiation models are commonly used to describe multi-wavelength data and explore the physical properties of high-energy sources, such as active galactic nuclei. However, these models often require a large number of free parameters.}
   {In the context of possible parameter degeneracy and the complex landscape of the parameter space, we study how the choice of the fitting procedure impacts the characterization of the source properties. Furthermore, we examine how the data coverage and the uncertainties associated to the data influence the model parameter degeneracy.}
   {We generated simulated spectral energy distribution datasets with different properties, which we then fit with a numerical model. The model describes the relevant radiation processes with seven free parameters. We compare different optimization algorithms and study the parameter degeneracy using t-distributed stochastic neighbor embedding. 
   
   Additionally, we applied the same fitting procedures to the observational data of two sources, Mrk 501 and PKS 0735+178.}
   {We demonstrate significant degeneracies in the seven-dimensional parameter space of the one-zone leptonic models caused by the incomplete wavelength coverage of the data. Given the same goodness-of-fit function, the best-fit result depends on the choice of the minimization algorithm.}
   {Source properties extracted from the best-fit solution to realistic datasets cannot be interpreted as the only solution due to significant degeneracies of the model parameters. Adding new energy ranges (e.g. MeV) and regular source monitoring would allow to reduce gaps in the data and significantly decrease the parameter degeneracy. }

   \keywords{Radiation mechanisms: non-thermal --
                Methods: numerical --
                Galaxies: BL Lacertae objects: general
               }

   \maketitle
%

\section{Introduction}

Blazars are a rare subclass of active galactic nuclei (AGNs) that shoot a relativistic jet close to the observer's line of sight \citep{1995PASP..107..803U}. Small viewing angles and the relativistic speed of the plasma in the jet make the Doppler boosting especially efficient, leading to high luminosity of blazar jets. The emission of blazars has a non-thermal nature and spans from radio frequencies to gamma rays. 

Blazars are the dominant sources of extragalactic GeV and TeV gamma rays \citep{2014JCAP...11..021D,2015ApJ...800L..27A, 2020ApJS..247...33A, 2022EPJST.231...27B}. 
They are also suggested as neutrino source candidates \citep[see][for a recent review]{2021Univ....7..492G} and candidate neutrino blazar associations have been identified~\citep{2018Sci...361..147I,2018Sci...361.1378I,2016NatPh..12..807K, 2017ApJ...843..109G, 2019ApJ...880..103G, Franckowiak_2020, Rodrigues_2021, 10.1093/mnras/stac3607}.

A good understanding of the blazar emission mechanisms is crucial for studying the nature of these sources and is essential for both gamma-ray and neutrino astronomy.

The spectral energy distribution of blazars exhibits a typical two-bump structure. It is commonly accepted that the low energy emission (from radio to UV or X-rays) is the synchrotron emission of relativistic electrons in the jet. The nature of the high-energy emission (X-ray to gamma rays) remains unclear. The simplest way to explain the blazar spectral energy distribution (SED) is to assume that all radiation originates from one emission region called ``blob''. Depending on the particle species that produce the high-energy emission, the radiation models can be divided into leptonic, hadronic, or leptohadronic. In one-zone leptonic models, only relativistic electrons accelerated in the jet are responsible for the observed SED. The high-energy emission is explained by Compton scattering of low-energy photons by the same electrons that produce the synchrotron emission at lower energies. Those models are called synchrotron self-Compton (SSC) models.
Purely hadronic models assume that gamma rays originate exclusively from proton synchrotron emission of ultra-relativistic protons, while the high-energy electron population is responsible for the low-energy emission. However, due to high proton energy requirements, the hadronic component can only be subdominant in blazars \citep{2020ApJ...893L..20L}. Purely hadronic models would also require a different accretion paradigm, which makes them highly disfavored \citep{2015MNRAS.450L..21Z}.
In leptohadronic models, both protons and electrons contribute to the high-energy part of the SED. By interacting with synchrotron photons, the protons produce hadronic cascades that contribute to the high-energy emission alongside the leptonic inverse Compton effect. 

Numerical modeling of the SEDs is applied to explain the properties of gamma-ray-emitting blazars \citep[][and many others]{2013ApJ...768...54B, 2013ApJ...771L...4C, 2014A&A...562A..12P, 2014ApJ...782...82D, 2015MNRAS.447...36P, 2015MNRAS.448..910C, 2015ApJ...804...74P} or for studying their neutrino emission \citep[][and many others]{2015MNRAS.448.2412P, 2018ApJ...864...84K, 2017ApJ...843..109G, 2020ApJ...891..115P, Rodrigues_2021, 2022MNRAS.509.2102G,2023MNRAS.519.1396S}. As many of these works note, complex radiation models typically have a high number of free parameters and require a lot of computational efforts (especially for time-dependent models) to properly fit the data. Often, models are fitted to the data based on a small number of probed parameter values or with several parameters fixed.  

With the current advancements in numerical modeling, a new generation of codes was developed. Software packages such as AM$^3$ \citep{2023arXiv231213371K}, LeHaMoC \citep{2024A&A...683A.225S}, or SOPRANO \citep{2022MNRAS.509.2102G} allow to perform fast computations of blazar SEDs. This, in turn, opens the possibility of performing data fitting and goodness-of-fit estimations with a high number of free model parameters. In fact, a high number of simulated models was already used for fitting observed blazar SEDs in \cite{ Rodrigues_2021, 2024A&A...681A.119R, 2024A&A...689A.147R, 2024arXiv240904165O} and for training neural networks in \cite{2024A&A...683A.185T,2024ApJ...963...71B,2024ApJ...971...70S}.

This paper addresses the challenges of high-level data fitting using radiation models. We highlight that the parameters of one-zone leptonic models are degenerate (see Section \ref{sec:smoothness}). In a thorough comparison of different optimization algorithms, we show that the parameter degeneracy leads to ambiguity in the data fitting and physical interpretation of the models. While this work considers only blazar one-zone leptonic radiation models, similar problems arise in the radiation models applied to other classes of sources, including Seyfert galaxies, gamma-ray bursts (GRBs), tidal disruption events (TDEs), and others. 

This paper is structured as follows. Section \ref{sec:setup} describes the physical setup of one-zone leptonic models in detail. Section \ref{sec:tsne} introduces the chosen methods for visualization of the multi-dimensional data. In Section \ref{sec:smoothness}, we demonstrate the irregularity of the parameter space due to the nature of the radiative models and discuss its implications.  Section \ref{sec:optimization} provides a comparison of the performance of different optimization algorithms tested on three sets of simulated data which are then applied to observational data in Section \ref{sec:real_data}. We discuss our results in the context of current state-of-the-art numerical modeling and its applications in Section \ref{sec:discuss} and summarize our findings in Section \ref{sec:concl}. For the calculations in this paper, we adopt a flat $\Lambda$CDM cosmological model with parameters $H_0 = 70$ km s$^{-1}$ Mpc$^{-1}$ and $\Omega_{\textrm{m}} = 0.3$.

\section{Leptonic models} \label{sec:setup}

The leptonic SSC model is the way to explain the blazar multi-wavelength emission with the smallest number of free parameters. For the numerical modeling of the SSC scenario, we utilized the open-source time-dependent code AM$^3$ \citep{2023arXiv231213371K}. AM$^3$ numerically solves the system of coupled integro-differential equations that describe the evolution of the particle spectra in a fully self-consistent manner. 

We assumed that electrons are accelerated to a simple power-law spectrum\footnote{Parameters with or without prime refer to the values in the jet or observer's frame, respectively.} $dN/d{\gamma'}_\mathrm{e} \propto {\gamma'}_\mathrm{e}^{-\alpha_\mathrm{e}}$ with spectral index $\alpha_\mathrm{e}$, spanning a range of Lorentz factors from  ${\gamma'}_\mathrm{e}^\mathrm{min}$ to ${\gamma'}_\mathrm{e}^\mathrm{max}$. While more complex electron spectra are possible, we chose the simplest physically motivated spectrum for a minimalistic setup. The energy spectrum of the electrons is normalized to the total electron luminosity parameter, $L'_{\mathrm{e}}$. These particles are injected into a single spherical blob of size $R_{\textrm{b}}'$ (in the comoving frame of the jet) moving along the jet with Lorentz factor $\Gamma$, where they encounter a homogeneous and isotropic magnetic field of strength $B'$. We assumed the jet is observed at an angle $\theta_{\textrm{obs}} = 1/\Gamma_{\textrm{b}}$ relative to its axis, resulting in a Doppler factor of $\delta_{\textrm{D}}=\Gamma_{\textrm{b}}$. We adopted a steady-state approximation to obtain the particle spectra. The characteristic escape time is set equal to the light-crossing time for all particles. We evolved the kinetic equations over several escape timescales to ensure that the steady state is reached. During photon propagation from the source to the observer, a part of the high-energy gamma rays is attenuated due to interactions with the extragalactic background light (EBL). Unless a different model is explicitly stated, this effect is accounted for in all models based on \cite{2011MNRAS.410.2556D}. The parameter space of the leptonic models is represented by seven free parameters summarized in Table \ref{tab:lep_pars}.

\begin{table}
    \centering
    \caption{List of leptonic model parameters.}
    \begin{tabular}{ll}
    \toprule
    Parameter & Description\\
    \midrule
    $R^\prime_\mathrm{blob}$, cm & Radius of the sperical emission zone region\\
    $B^\prime$, G & Strength of the homogeneous magnetic field\\
    $\Gamma_\mathrm{b}$ & Blob Lorentz factor\\
    $\gamma_\mathrm{e}^{\prime\mathrm{min}}$ & Minimal electron Lorentz factor\\
    $\gamma_\mathrm{e}^{\prime\mathrm{max}}$ & Maximal electron Lorentz factor\\
    $\alpha_\mathrm{e}$ & Power-law index of the electron energy distribtion\\
    $L^\prime_\mathrm{e}$, erg s$^{-1}$ & Total electron luminosity\\
    \bottomrule
    \end{tabular}
    
    \label{tab:lep_pars}
\end{table}

\section{Visualization of parameter spaces} \label{sec:tsne}

A typical procedure of data fitting is the optimization of a goodness of fit. As a goodness-of-fit function, we choose a reduced $\chi^2$-function (i.e. divided per number of degrees of freedom) defined as
  \begin{equation} \label{eq:chi2}
 \chi^2(\theta)/\textrm{n.d.f.~} =~\frac{1}{N-N_\mathrm{par}+1} \sum\limits_i\frac{\bigl(F_i^\mathrm{data}-F_i^\mathrm{model}(\theta)\bigr)^2 }{\sigma_i^2} \,,
\end{equation}  
\noindent
where $N$ is the number of data points, $N_{\mathrm{par}}=7$ the number of free parameters in the model, $F^\mathrm{data}$ are the observed fluxes, $F^\mathrm{model}$ are the predicted fluxes by the model, and $i$ is the summation index that corresponds to the observed frequency values, and $\sigma_i$ are the flux measurement uncertainties. The function input $\theta\in\mathbb{R}^7$ are the model parameters from Table \ref{tab:lep_pars}. While in the general case different goodness-of-fit functions are possible, we select the reduced $\chi^2$ function commonly used in physics and astronomy as its values directly indicate a poor fit ($\chi^2(\theta)/\textrm{n.d.f.~}{\gg}1$), an overfitting ($\chi^2(\theta)/\textrm{n.d.f.~}{\ll}1$) or a good fit ($\chi^2(\theta)/\textrm{n.d.f.~}{\approx}1$). The model that produces the minimal value of the reduced $\chi^2$ is defined as the best fit. If the SED contains upper limits on the flux values, they do not contribute to the reduced $\chi^2$ value as long as the model predictions are below their values. In the opposite case, a large value ($10^5$) is added to the reduced $\chi^2$ so that a model overshooting upper limits is considered highly unfavorable.

To compare the results of the selected minimization procedures (Section \ref{sec:optimization}), it is essential to define how close or how far multidimensional vectors of the model parameters are located with respect to each other in the parameter space depending on their goodness-of-fit value.
Visualization of the parameter space reveals multiple regions with comparable $\chi^2$ values. For this purpose, we used t-distributed stochastic neighbor embedding (t-SNE), a machine learning technique for mapping high-dimensional data into low-dimensional representations \citep{van2008visualizing}. The dimensionality reduction in t-SNE is performed by matching the distribution of point similarities (defined as probabilities non-linearly proportional to the metric distances) in high-dimensional space and low-dimensional space of two abstract coordinates. As a result, points with higher similarities are arranged closer to each other in the 2-dimensional map but the ratio between the distances of points in high-dimensional space and in low-dimensional space is not conserved. An important concern of t-SNE usage is the discrimination of truely close points and artificial clustering. A hyperparameter that can affect artificial clustering is called perplexity. It reflects the amount of nearest neighbors and balances the analysis of local and global data features. The details on hyperparameter turning for ensuring the adequate reflection of point similarities can be found in the Appendix \ref{sec:appendix:hyperparameters}. 

\begin{figure} [htbp!]
    \centering
    \includegraphics[width=0.9\linewidth]{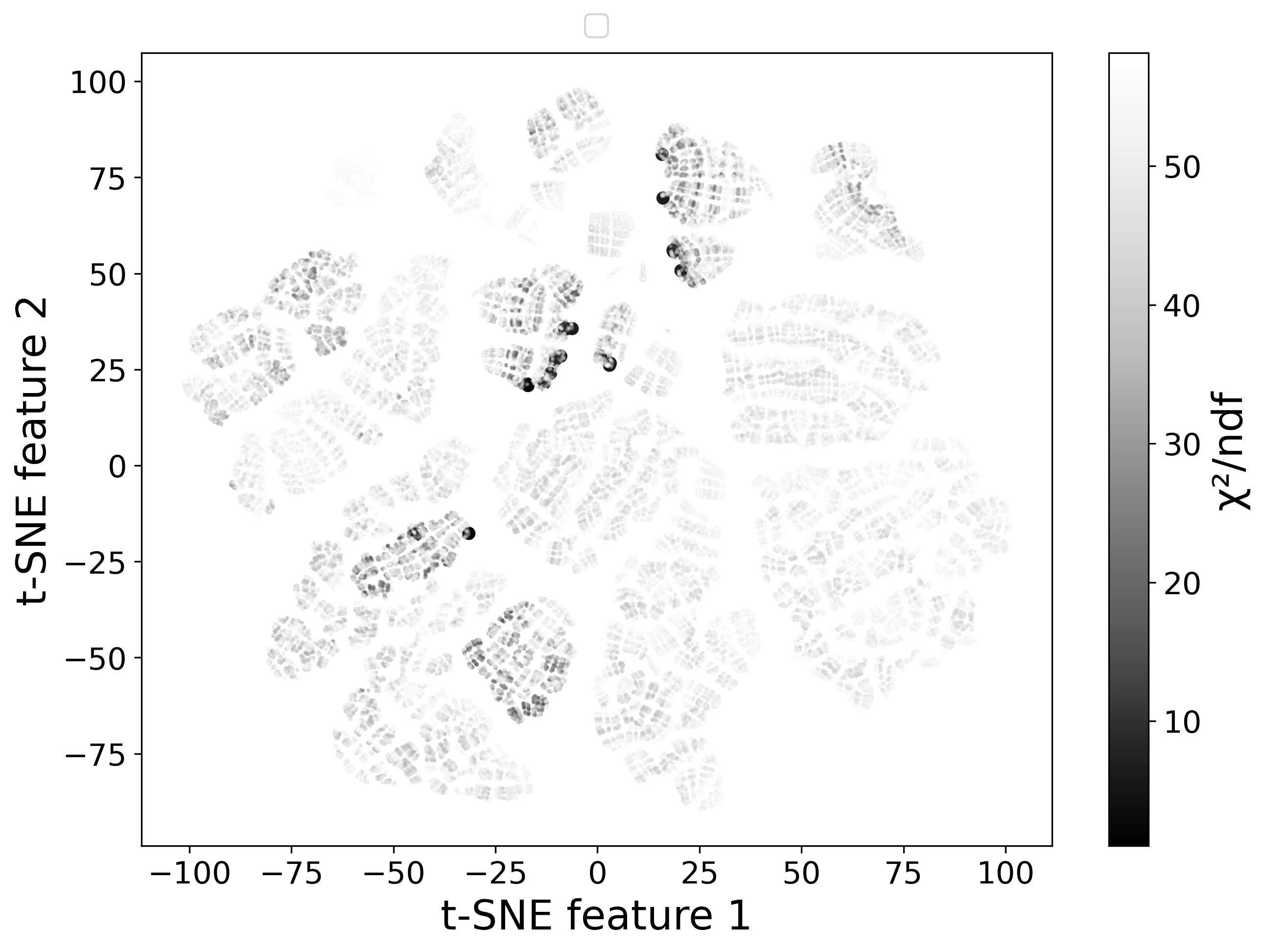}
    \caption{Example of the seven-dimensional parameter space mapping with t-SNE (grid scan results for simulated dataset 2, see Sec \ref{sec:simdata}). Each point represents one model consisting of seven parameters. Darker colors correspond to lower $\chi^{2}/$n.d.f. values. The plot shows the first 50000 points with the smallest $\chi^2$/n.d.f. value, and the perplexity is set to 30.}
    \label{fig:example_tsne}
\end{figure}

In this paper, we used the t-SNE tool from the \texttt{scikit-learn} open-source Python library \citep{scikit-learn}. Before normalizing the values, we transformed those parameters with values spanning over several magnitudes ($R'_\mathrm{blob}$,  $\gamma'_\mathrm{min}$, $\gamma'_\mathrm{max}$ and $L'_\mathrm{e}$) to logarithmic scale. This step should be performed in addition to normalization to ensure a more balanced contribution from all parameters by avoiding skewed distributions during normalization to a [0,1] scale. Without this step, extreme values (e.g. $10^{40}$ vs. $ 10^{47}$) were dominant in the analysis, leading to neglection of smaller but meaningful variations.

Then the low-dimensional map of the parameter space was built using t-SNE. The goodness-of-fit value that corresponds to each parameter vector is used as a color scale to highlight the regions in the parameter space with the best solutions. Fig.~\ref{fig:example_tsne} shows an example of such a t-SNE output which represents the parameter space of the leptonic models applied to the simulated dataset 2 (see Sec. \ref{sec:simdata}).

The darker regions in the plot correspond to the regions where good-fit solutions can be found. The plot shows 50\,000 models with different combinations of model parameters. Fig.~\ref{fig:example_tsne} highlights five regions where the solutions with low reduced $\chi^2$ < 7 are located. Those solutions are shown with black markers bigger than the rest of the points. Due to the systematic and statistical uncertainties of the data, slight variations of the model parameters can still produce an SED with a small goodness-of-fit value. Therefore, the solutions are expected to form island-like structures with close but slightly deviating values of model parameters. Apart from this, there might be a completely different solution from the distant region in the parameter space that produces a similarly good fit. This can be seen in Fig.~\ref{fig:example_tsne} where several distant regions contain good solutions which, in turn, are surrounded by candidates with slightly worse goodness of fit. For the following comparisons, the close location of two points in the t-SNE plots should be interpreted as close parameter values for the corresponding models. On the contrary, points scattered over the t-SNE output are located in distant regions and correspond to physically different solutions.

\section{Challenges for parameter fitting} \label{sec:smoothness}

Minimizing the goodness-of-fit function is challenging because the function is expected to be neither convex\footnote{A function $f: \mathbb{R}^n \rightarrow \mathbb{R}$ is convex if for all $0\leq t \leq 1$ and all $x, y \in \mathbb{R}^n$ $f(tx + (1-t)y)\leq tf(x) +(1-t)f(y)$.} nor smooth\footnote{All first derivatives are continuous.}. The first reason lies in the nature of the radiation processes. Processes that have thresholds (like $\gamma \gamma \rightarrow e^+ e^-$) or show very different features at different energies (like the Thomson and Klein-Nishina regimes in the case of the inverse Compton effect) impact the smoothness of the goodness-of-fit function since even a small change in some of the parameters leads to a drastically different shape, meaning that the goodness-of-fit function cannot be differentiated numerically.
The second reason for expecting a non-smooth and non-convex goodness-of-fit function is the limited amount of available photon flux measurements. In the ideal case, for data covering many orders of magnitudes without gaps, the shape of the SED is unambiguously defined by the spectrum of the electron population and the conditions in the emission zone. In reality, however, the SED almost always has gaps in some energy ranges due to the limited number of instruments covering only selected energy ranges. Different electron populations can produce similar photon-flux levels with differences that could be spotted only in the missing data regions. By varying, for example, the electron maximum Lorentz factor, spectral index, and electron luminosity (which normalizes the spectrum), similar levels of photon fluxes can be achieved for different combinations of these parameters. Additionally, the observed luminosity follows $L \simeq \delta_{D}^4 L'$, meaning that for each value of electron luminosity in the source, there exists such a bulk Lorentz factor that the observed luminosity remains unchanged. However, different blob Lorentz factors would produce different SED shapes. This could be corrected by variations in the electron spectral parameters, thus resulting in very similar SEDs. The existence of multiple good-fitting solutions suggests that any goodness-of-fit function is expected to be non-convex. Hence, formally defining optimal fits is straightforward, but computing them is challenging. We will address this topic in Section \ref{sec:optimization}.

\section{Comparison of minimization procedures using simulated data} \label{sec:optimization}

In a non-smooth and non-convex parameter space, the choice of the minimization algorithm can impact the result of the fitting. This also leads to implications on the physical interpretation of the models. We selected five minimization techniques most commonly used in the literature: \texttt{Minuit}, grid scan, two evolutionary algorithms, and Markov Chain Monte Carlo. In this Section, we test the performance of the selected algorithm on simulated pseudo-data, with known true values of the model parameters. To study the effects of wavelength coverage of the data and of the size of the uncertainties associated to the data, we created three simulated datasets. We compare the best-fit results of each algorithm with the true values.

\subsection{Simulated data} \label{sec:simdata}

\begin{figure*}[htbp!]

\centering
     \begin{subfigure}{0.33\textwidth}
         \centering
         \includegraphics[width=\textwidth]{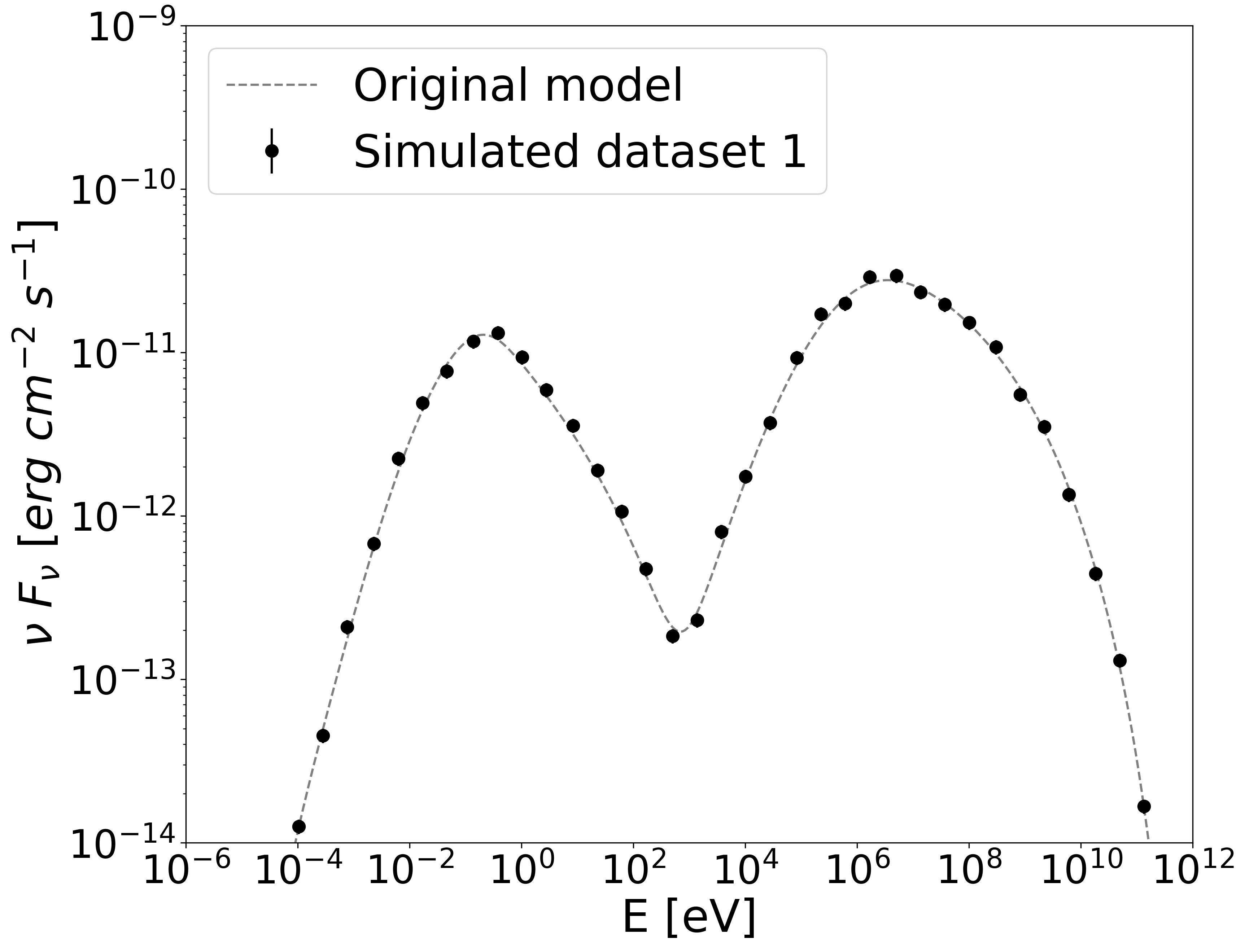}
         \label{fig:y equals x}
     \end{subfigure}
     \begin{subfigure}{0.33\textwidth}
         \centering
         \includegraphics[width=\textwidth]{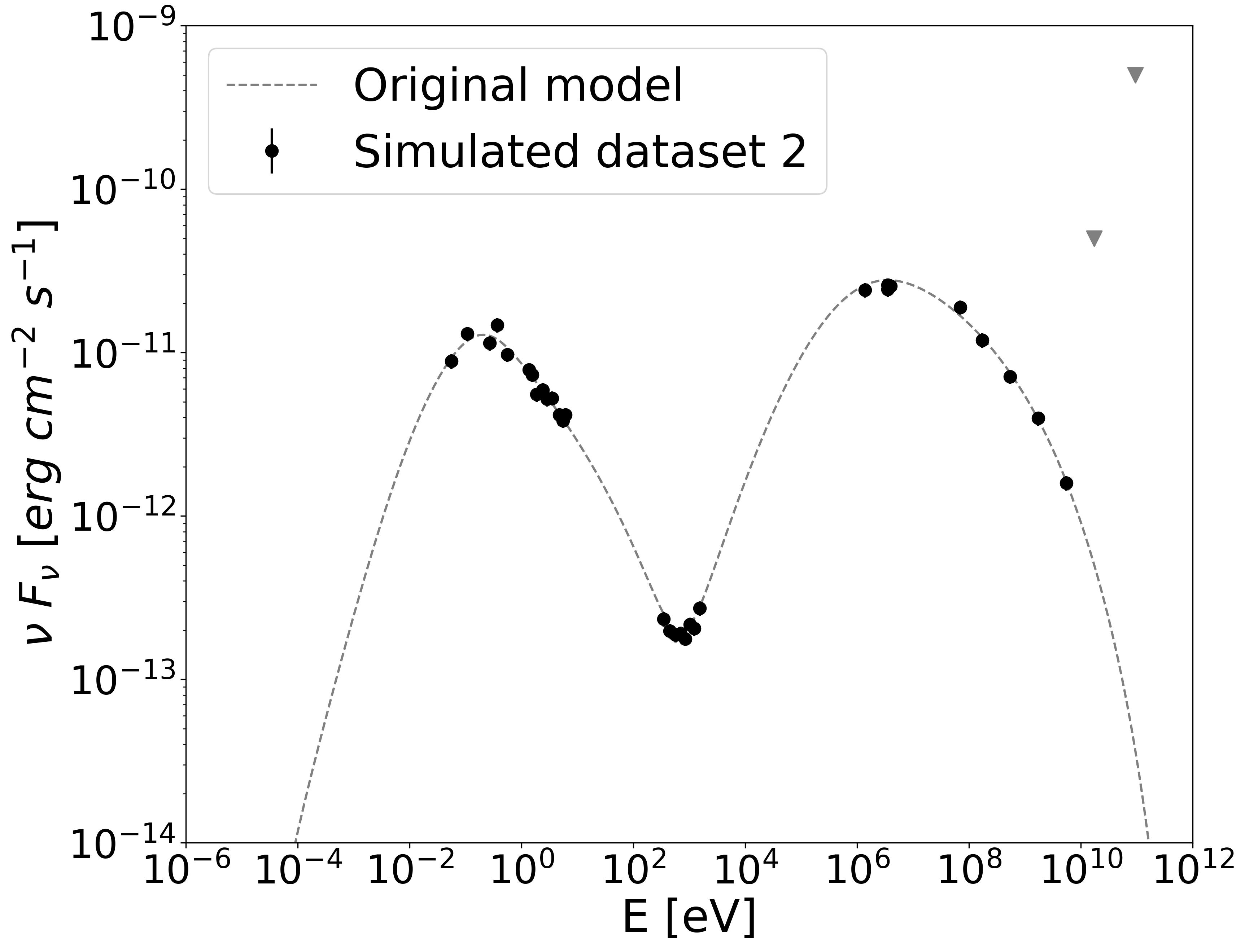}
         \label{fig:three sin x}
     \end{subfigure}
     \begin{subfigure}{0.33\textwidth}
         \centering
         \includegraphics[width=\textwidth]{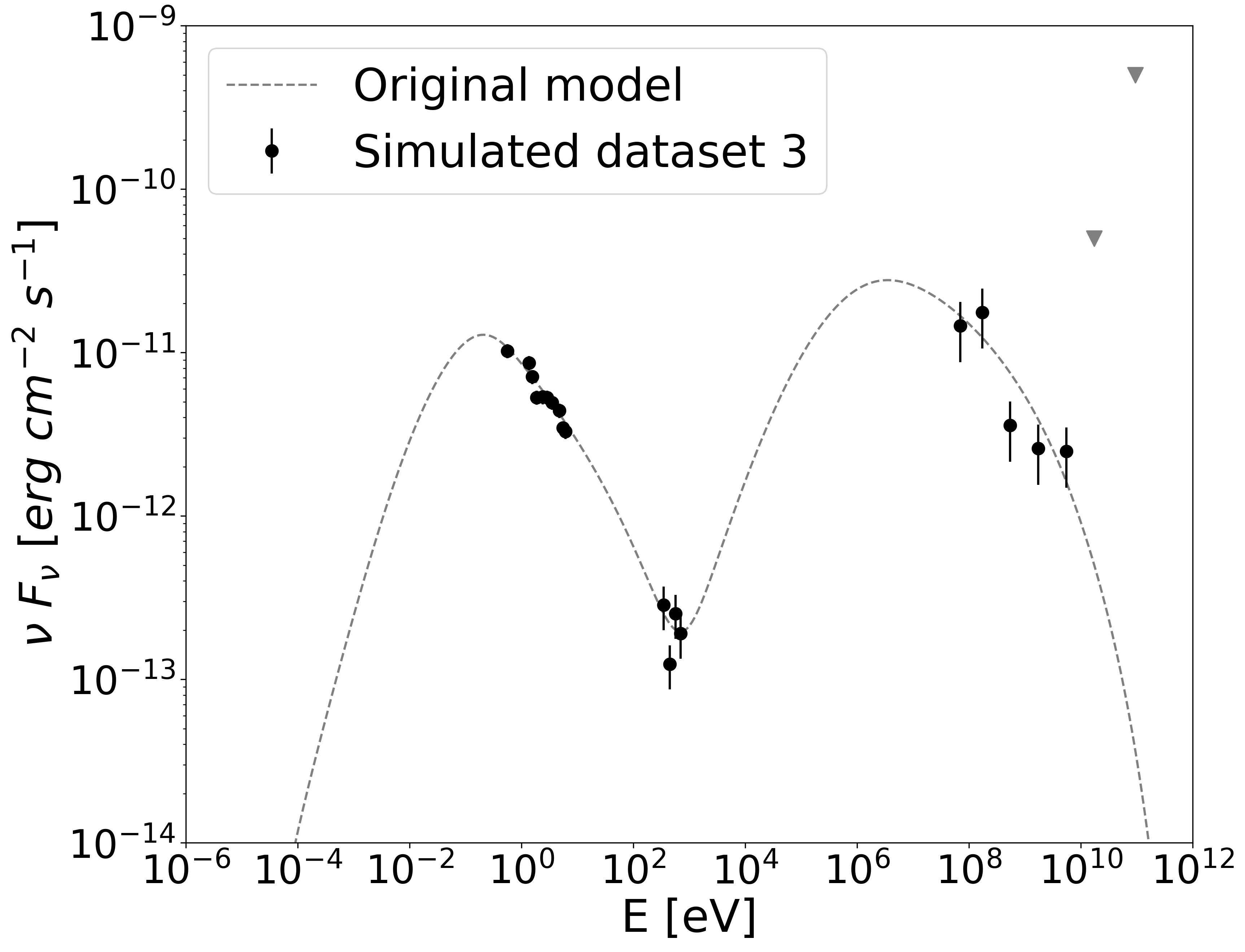}
         \label{fig:five over x}
     \end{subfigure}
        \caption{Simulated datasets. The black dots show the data points in the simulated datasets. The dashed curve shows the original model that was used to generate the data. The grey triangles in the last two plots are upper limits.}
        \label{fig:Simulated_datasets}

\end{figure*}

Unlike real blazars, where the true underlying physical parameters are unknown, fitting pseudo-data with known parameters allows for a direct comparison of the accuracy of different minimization algorithms.  
The simulated data was generated with AM$^{3}$ using the parameters from the first row of Table \ref{table:table_simdata}. To create a SED in a realistic flux and energy range, we choose parameter values near the best-fit leptonic model parameters for the quiescent state of the blazar PKS~0735+178 provided in \citet{2024arXiv240904165O}. Based on this SED, we created three datasets, each resembling a different wavelength coverage of the data and different uncertainties associated to the flux measurements.

For dataset 1, we equally spaced the data points on a logarithmic scale across all energies. To mimic the uncertainty of the flux measurements, we added a random value distributed normally around zero and with a standard deviation corresponding to 10 \% of the flux value. After the new, shifted fluxes are calculated, we assume a 10\% uncertainty for all data points. Dataset 1 represents the case of a ``perfect'' SED.

Dataset 2 resembles a more realistic case with good data coverage. We selected energies that correspond to K, z, I, r, g filters in the optical range and $U$, $B$, $W1$, $M2$, and $W2$ corresponding to the ultraviolet (UV) bands of the UVOT instrument onboard the \textit{Neil Gehrels Swift} Observatory \citep{2005SSRv..120...95R}, a typical energy range of X-ray spectra from \textit{Swift}-XRT \citep{2004SPIE.5165..201B} and NuSTAR \citep{2013ApJ...770..103H}, the centers of the four COMPTEL \citep{1993ApJS...86..657S} energy bands (0.75--1, 1--3, 3--10 and 10--30 MeV), and typical energy bands of the gamma-ray telescope \textit{Fermi}-LAT \citep{Atwood_2009}. The two highest energy data points represent sensitivity-based upper limits. The uncertainty of the flux measurements is added in the same way as it was done for dataset 1.

Dataset 3 represents a more pessimistic case with a limited amount of available data. The data covers only the optical/UV, the typical \textit{Swift}-XRT bands in X-rays, and the \textit{Fermi}-LAT gamma-ray range. Unlike the two previous datasets, we assumed different flux uncertainties for different energy bands: 10\% in optical/UV, 30\% in X-rays, and 40\% in gamma rays. The procedure for accounting for the uncertainties remained the same as for dataset 1. Thus, taken together, the three datasets provide a panorama of possible situations.

\subsection{Tested optimization algorithms}

All selected algorithms aim to minimize the same reduced $\chi^2$ function defined in Eq.~\eqref{eq:chi2}. For the simulated data, they search for the best solutions within the same parameter space limited by the boundaries shown in Table \ref{tab:lep_pars_bound}.

\subsubsection{Grid scan}
A grid scan is an approach where values of the goodness-of-fit function are evaluated for each combination of the parameters equally spaced between defined boundaries (Table \ref{tab:lep_pars_bound}). This simple setup allowed us to cover the whole parameter space homogeneously and highlights the regions of interest corresponding to low reduced $\chi^2$ values. We selected ten points per parameter, probing in total $10^7$ points in the parameter space. Usually, the discretization of the grid is not good enough to find a precise best-fit solution. Therefore, as a second step, we locally minimized the best result from the grid scan using \texttt{Minuit} (using \texttt{simplex} followed by \texttt{migrad}, see Sec. \ref{sec:minuit}). Since all neighboring points of the current best-fit grid scan solution have a higher reduced $\chi^2$ value, we adopted their values as new boundaries for searching for the new best fit. The local minimization step required an additional probing of 300 points.

\subsubsection{Minuit} \label{sec:minuit}

\texttt{Minuit} is a software library for numerical minimization developed at CERN \citep{james1998minuit}. For this work, we used the Python interface for \texttt{Minuit} called \texttt{iminuit} \citep{iminuit}. We start the minimization with the Nelder-Mead simplex method \citep{nelder1965simplex} due to its robustness. The independence of the gradient makes it especially suitable for the large parameter space of the leptonic models. Due to the nature of the radiation processes, small changes in the model parameters may induce such a change in the reduced $\chi^2$ that it becomes numerically non-differentiable at this point. After the \texttt{simplex} method, we call \texttt{migrad}, a method based on gradient descent, to refine the best-fit solution locally. 

We initialized the simplex algorithm based on some physically motivated assumptions on the model parameters derived from the SED features. We set the maximum number of function calls to 1200. Since the first run never yielded an acceptable value of reduced $\chi^2$, we repeated the procedure while refining the boundaries of the parameters and reducing the parameter space around the previously found region. Our new initial point in the second run was set as the best-fit solution from the previous run. The second run required around $300$ function calls.

\subsubsection{Genetic algorithm}

Another minimization technique is the genetic algorithm \citep{kramer2017genetic}. Inspired by biological evolution, a set of candidate solutions is evaluated and evolved toward an optimal solution. We used the evolutionary functions from the DEAP module \citep{DEAP_JMLR2012} in Python. The algorithm starts with initializing a population of random parameter vectors. The fitness of each individual is evaluated by calculating each $\chi^{2}/$ndf value for a given dataset. Parents with better fitness values are selected to generate the offsprings via crossover and mutation. Crossover imitates the combination of DNA sequences during reproduction. The crossover operators mix genetic material of parent solutions, and there are various approaches to achieve this. In this work, the method involves splitting the two parent solutions and reassembling them alternately. For mutation, certain properties of the selected solutions are randomly altered to generate new offspring genes. Another parameter is the independent probability for each parameter to mutate. The crossover, mutation, and independent probability are controllable parameters of the genetic algorithm. Their values used for this work are given in Table \ref{tab:GA_params}. After the variation with crossover and mutation, the previous parent population is replaced by the new offspring, and the process is repeated through several generations. We take the standard deviations of the parameter values in the last generation as the uncertainty in the parameter estimation. 
\begin{table}[H]
    \centering
    \caption{Parameters of the genetic algorithm}
    \begin{tabular}{lc}
    \toprule
    Parameter & Values \\
    \midrule
    Number of generations & 70 -- 80\\
    Number of individuals & $10^{4}$\\
    Mutation probability & 0.33\\
    Independent probability & 1/7\\
    Crossover probability & 0.20 \\

    \bottomrule
    \end{tabular}
    
    \label{tab:GA_params}
\end{table}

\subsubsection{CMA-ES}

The Covariance Matrix Adaption Evolution Strategy \citep[CMA-ES,][]{2016arXiv160400772H} is the second evolutionary algorithm that is used in this work. The procedure is similar to the genetic algorithm, with the main difference lying in the mutation treatment. The mutation strength indicates how strongly the offspring differs from the parent population. CMA-ES adjusts the mutation step size based on the success of the previous mutations, which allows the algorithm to learn the mutation strength during the evolution. Compared to the genetic algorithm, this strategy requires less manual parameter tuning. While the multivariate Gaussian distribution is the generalization of the Gaussian distribution to higher dimensions, the covariance matrix can be interpreted as the generalization of the variance. From a multivariate Gaussian distribution, the candidate solutions are generated and evaluated and the sampling distribution is updated for the next iteration to increase the likelihood of selecting better solutions. Further information about CMA-ES can be found in e.g. \cite{6790628}. We used the python package \texttt{pycma} by \citet{hansen2019pycma}. The initial step size for the first generation has to be set in the beginning. After the first generation, this quantity is adjusted in each generation. In this work, the initial step size was set to $\sigma$ = 2. The population size is fixed to 5000 individuals. Similar to the genetic algorithm, the uncertainty in the parameter estimation is assumed to be the standard deviations of the corresponding parameter values in the last generation.

\subsubsection{MCMC}
Markov chain Monte Carlo (MCMC) is a method of drawing samples from probability distributions. The posterior probability distribution allows to find the optimal values of the model
parameters. Despite the popularity of MCMC in data fitting, \cite{Hogg_2018} note that just searching the parameter space is not a good motivation for MCMC usage since it is primarily a sampler. Motivated by the common usage of MCMC in blazar modeling \citep[see e.g.][]{10.1093/pasj/psaa028, 2020ascl.soft09001T,   tzavellas2023applicationneuralnetworkssynchrocompton, Sciaccaluga_2024, Hervet_2024}, we investigate this method for data fitting as well.

In this work, we utilize the affine invariant Markov chain Monte Carlo ensemble sampler \citep{10.2140/camcos.2010.5.65} and the \texttt{emcee} package as its Python implementation \citep{emcee}. The emcee algorithm requires the user to define a log-likelihood function that evaluates the quality of the fit. Based on this function, it autonomously constructs the posterior probability density function, employing a specified number of steps, walkers, and burn-in samples. 

\cite{Hogg_2018} recommend selecting some sensible parameter values as a starting point, close to the optimum but not exactly the optimum. Following this, we initialize the walkers in the vicinity of the best-fit Minuit result by adding to the parameter values random values that follow standard normal distribution scaled by 10\% of the parameter value. 

We selected a ``flat'' (improper) prior for the model parameters with the boundaries of the distributions fixed to the values from Table \ref{tab:lep_pars_bound}. As for the log-likelihood function, we first defined it as $\ln (P) = - 1/2\, \log_{10}(\chi^2)$. The logarithmic scale of $\chi^2$ was chosen to overcome regions of very low probability that may separate the regions with high probability. However, the optimal parameters found in this MCMC run did not yield a satisfactory reduced $\chi^2$ value. We assume that the logarithmic reduced $\chi^2$ had the effect of losing precision near the best-fit solution. To exclude this effect, we repeat the MCMC (by performing run 2) with the new initial point as the previously found optimal solution (from run 1). 
To find a precise solution, we complete a second MCMC run with $\ln (P) = - 1/2 \, \chi^2$ while limiting the parameter space to the high-likelihood region identified during the first run.

\begin{table}[H]
    \centering
    \caption{Parameters of the MCMC runs}
    \begin{tabular}{lcc}
    \toprule
    Parameter & MCMC run 1 & MCMC run 2\\
    \midrule
    Log-likelihood function & $\ln (P) = - \frac{1}{2} \log_{10}(\chi^2)$ & $\ln (P) = - \frac{1}{2} \chi^2$\\
    Log-prior & flat & flat \\
    Number of walkers & 50 & 50\\
    Number of steps & 1000 & 200\\
    Burn-in samples & 100 & 20\\
    Initial point & from Minuit & from run 1\\

    \bottomrule
    \end{tabular}
    
    \label{tab:MCMC_params}
\end{table}

\subsection{Algorithm comparison}

With regard to the performance of the algorithms, the grid scan offers a decisive advantage as it can effectively show multiple regions with possible good-fit solutions. Since the other algorithms converge to a single best-fit model, the grid scan is useful for considering multiple solutions. It has no controllable algorithm parameters other than the boundaries of the parameter space and the choice of the discretization. Nonetheless, it is also the most computationally expensive of all methods. In the case of more complex models, the number of required generated models (i.e. probed points in the parameter space) grows faster than exponential. Additionally, the results depend on the choice of the step size and the boundaries. Some of the regions of interest may be lost between the grid points if the step between the grid points is too large. On the other hand, the grid scan results can be reused for other sources by just recalculating the $\chi^{2}$/n.d.f. for another dataset and followed by a computationally inexpensive local minimization. 

The genetic algorithm and CMA-ES, both from the family of evolutionary algorithms, have different performances. 
While the initial population is generated randomly (not user-defined) for both algorithms, the genetic algorithm has more controllable parameters. The CMA-ES is characterized by self-adaptivity and converges dramatically faster than the genetic algorithm. This can be shown with Figure \ref{fig:convergence}.

The Minuit approach (utilizing simplex and migrad) is the least computationally expensive method. It also requires only the parameter intervals and the initial steps to be defined by the user. Despite the need to run Minuit a couple of times in a ``nested'' way, each time refining the boundaries and increasing the precision, the final best-fit result has a goodness of fit similar to those from more computationally expensive algorithms. The outcome of the Minuit optimization in the parameter space of the one-zone radiation models depends on the choice of the initial point. While an ``educated'' guess based on the SED features may quickly lead to a good fit, random initialization may lead to local minima or even a failed fit. 

MCMC provides the posterior distributions of the model parameters which allow the characterization of the parameter space near the best fit. 
It requires assumptions on the parameter distribution in the absence of any data (log-prior), the number of walkers, and the number of steps. Similarly to Minuit, the performance of MCMC in the parameter space of the one-zone leptonic models depends on the choice of the initial point. If initialized near the known optimum, MCMC converges fast with the required number of generated models being similar to those from the evolutionary algorithms. If the selected initial point is far from the optimum, it may require the same or a larger number of generated models as the grid scan for convergence. The comparison of all selected algorithms is also summarized in Table \ref{tab:performance}.

\subsection{Results}
\begin{figure*}[htbp!]

\centering
      \begin{subfigure}{0.4\textwidth}
         \centering
         \includegraphics[height=6cm]{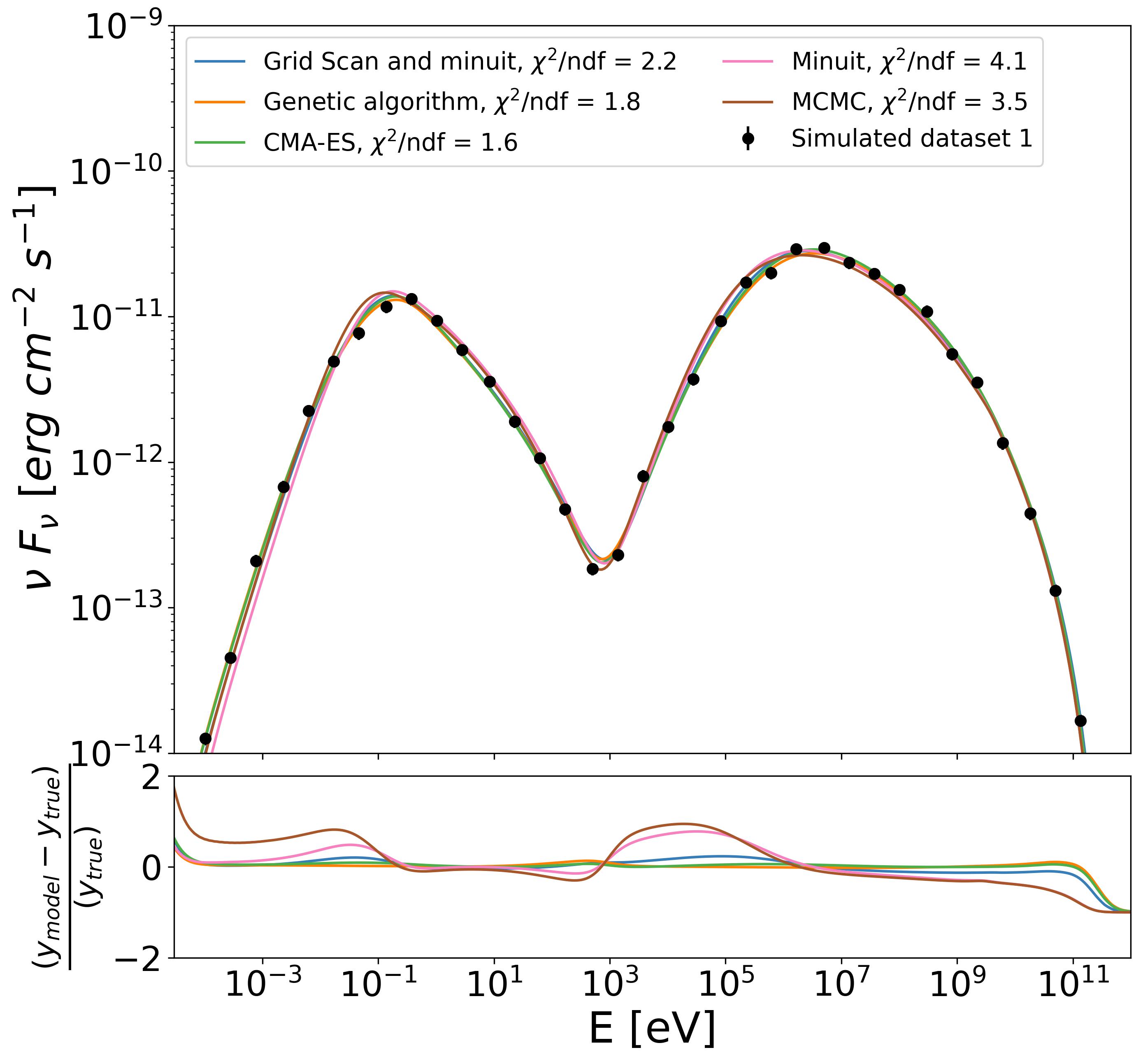}
         
         \label{fig:ds1_seds}
     \end{subfigure}
     \begin{subfigure}{0.4\textwidth}
         \centering
         \includegraphics[height=6cm]{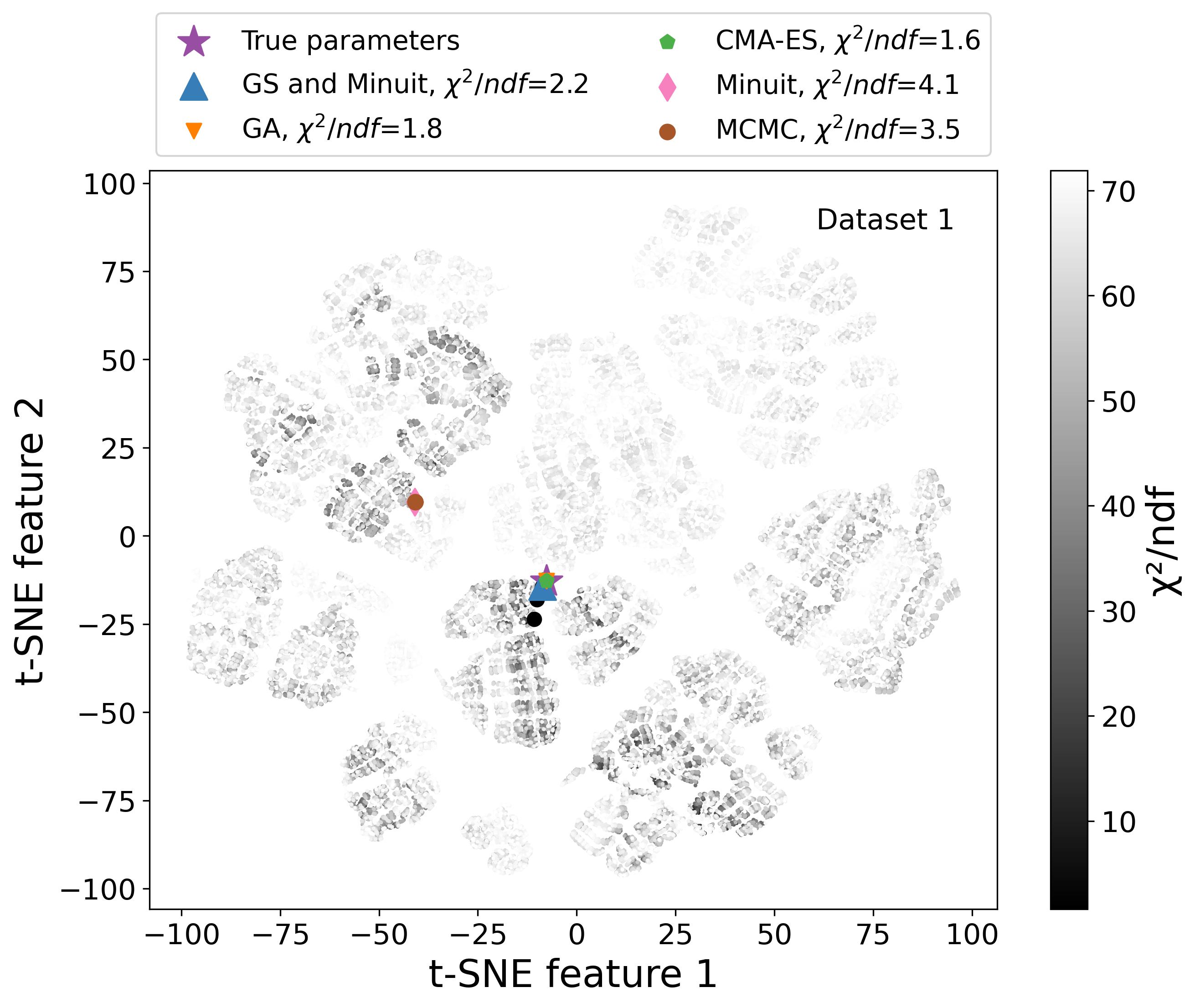}
         
         \label{fig:df1_tsne}
     \end{subfigure}
        \caption{Results of the SED fitting for dataset 1. Left panel, upper plot: the best-fit results from all selected optimization algorithms. Left panel, lower plot: relative deviation between the true values of $\nu F_\nu$ and those of the best-fit solutions. Right panel: the location of the best-fit solutions in the global parameter space shown in a t-SNE map.}
        \label{fig:results_simdata1}

\end{figure*}

\begin{figure*}[htbp!]

\centering
     \begin{subfigure}{0.4\textwidth}
         \centering
         \includegraphics[height=6cm]{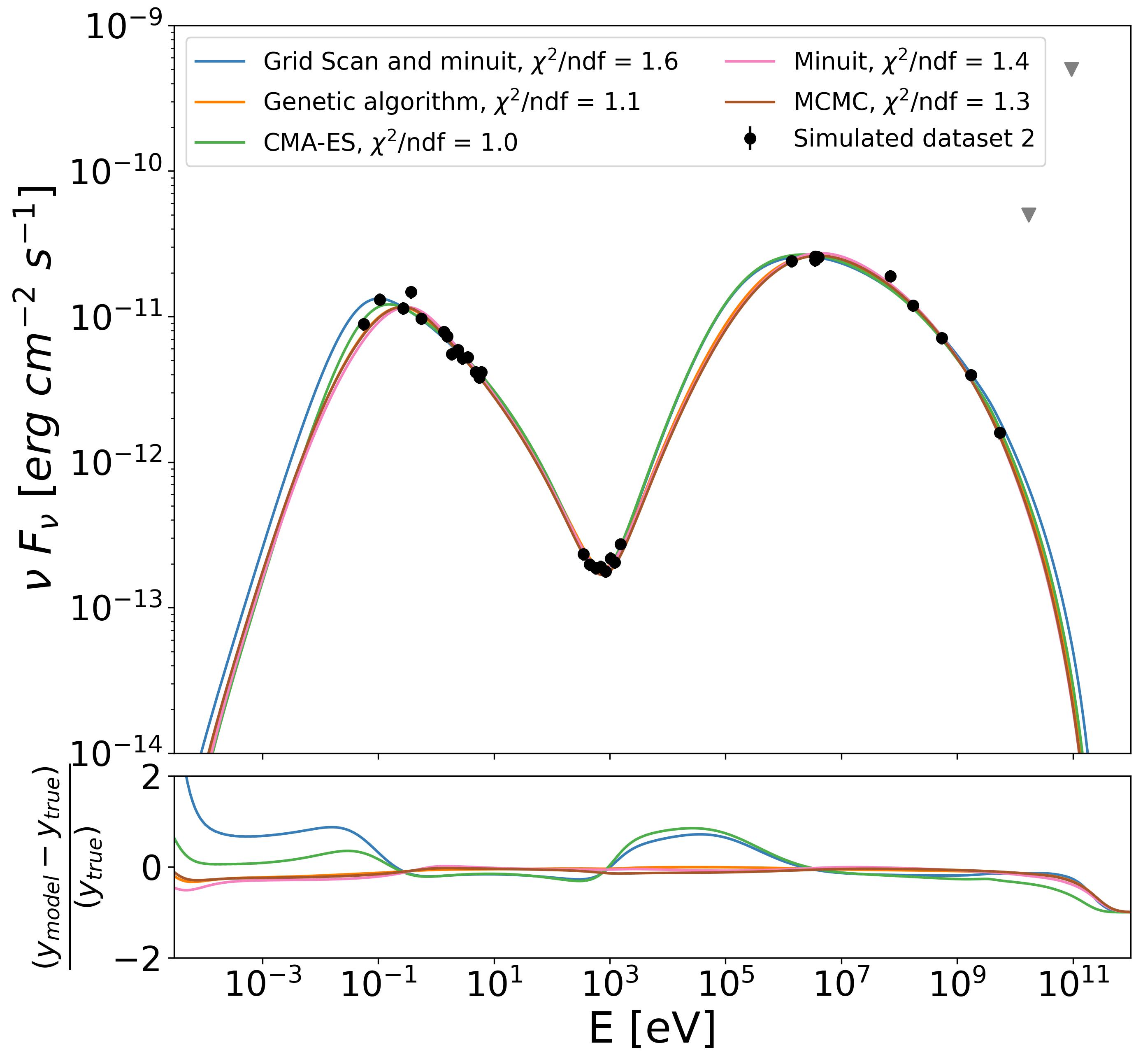}

     \end{subfigure}
     \bigskip
     \begin{subfigure}{0.4\textwidth}
         \centering
         \includegraphics[height=6cm]{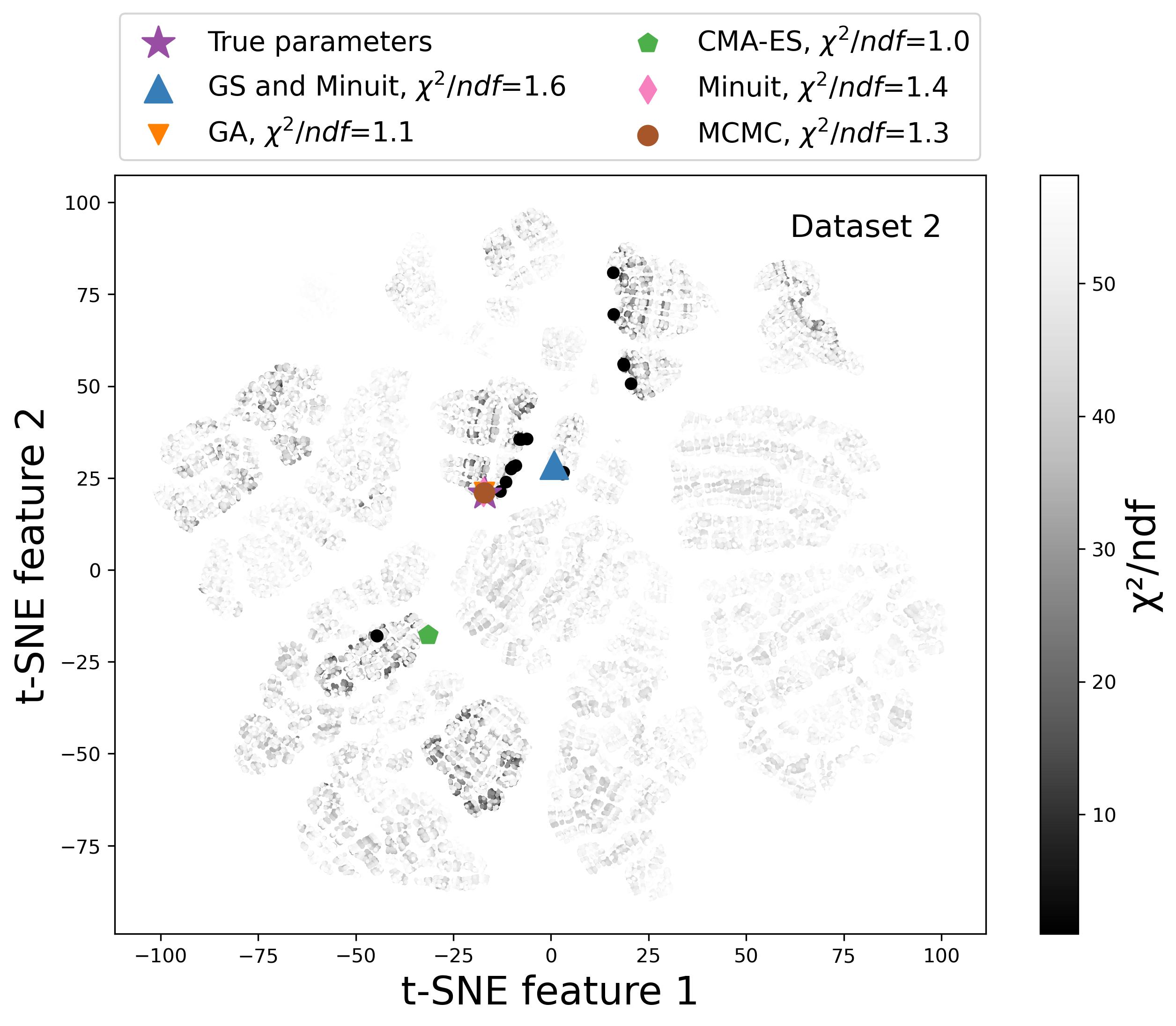}

     \end{subfigure}
     
        \caption{Same as Fig. \ref{fig:results_simdata1} but for dataset 2.}
        \label{fig:results_simdata2}

\end{figure*}

\begin{figure*}[htbp!]

\centering
     \begin{subfigure}{0.4\textwidth}
         \centering
         \includegraphics[height=6cm]{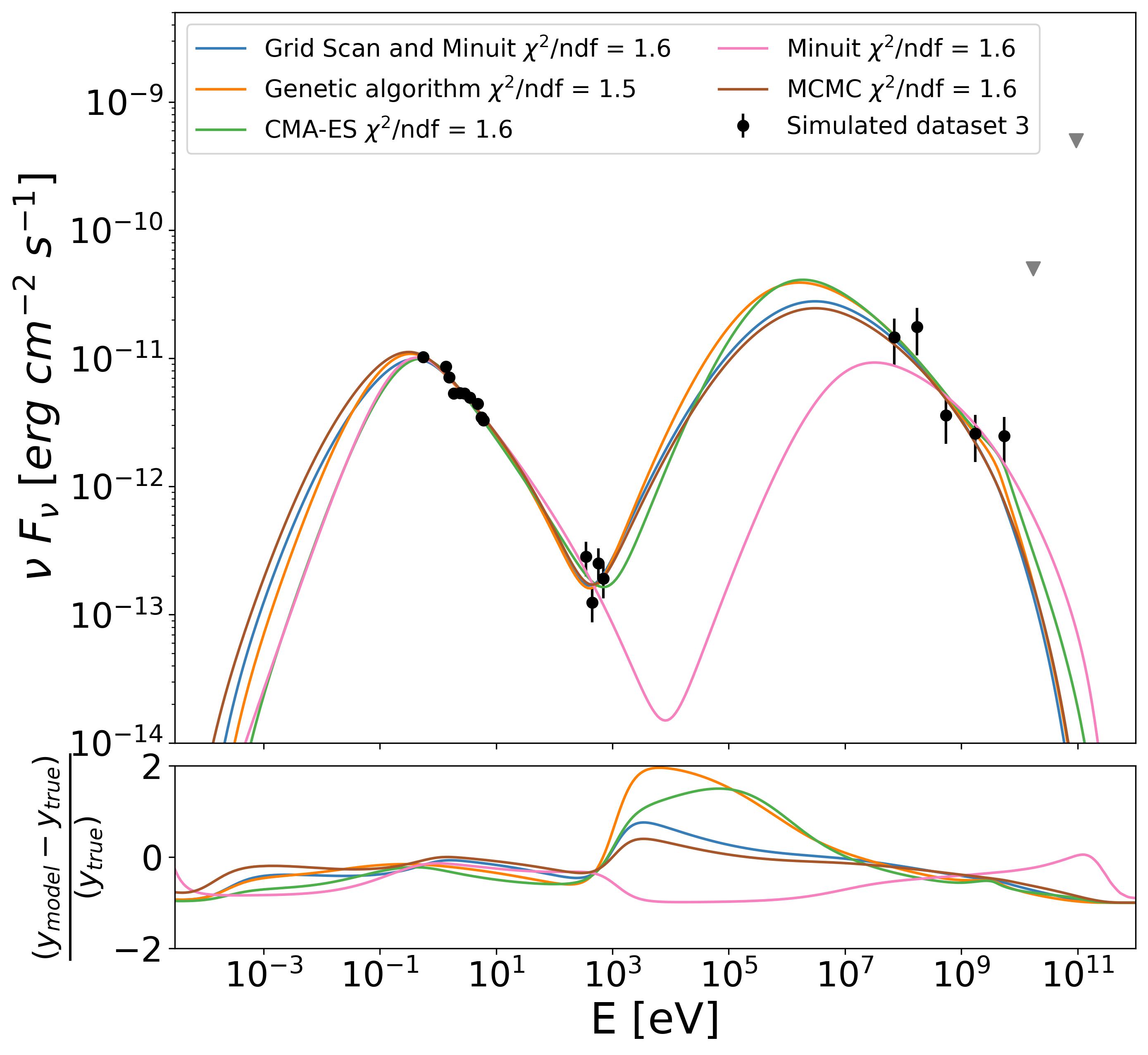}
         
         \label{fig:five over x}
     \end{subfigure}
     \begin{subfigure}{0.4\textwidth}
         \centering
         \includegraphics[height=6cm]{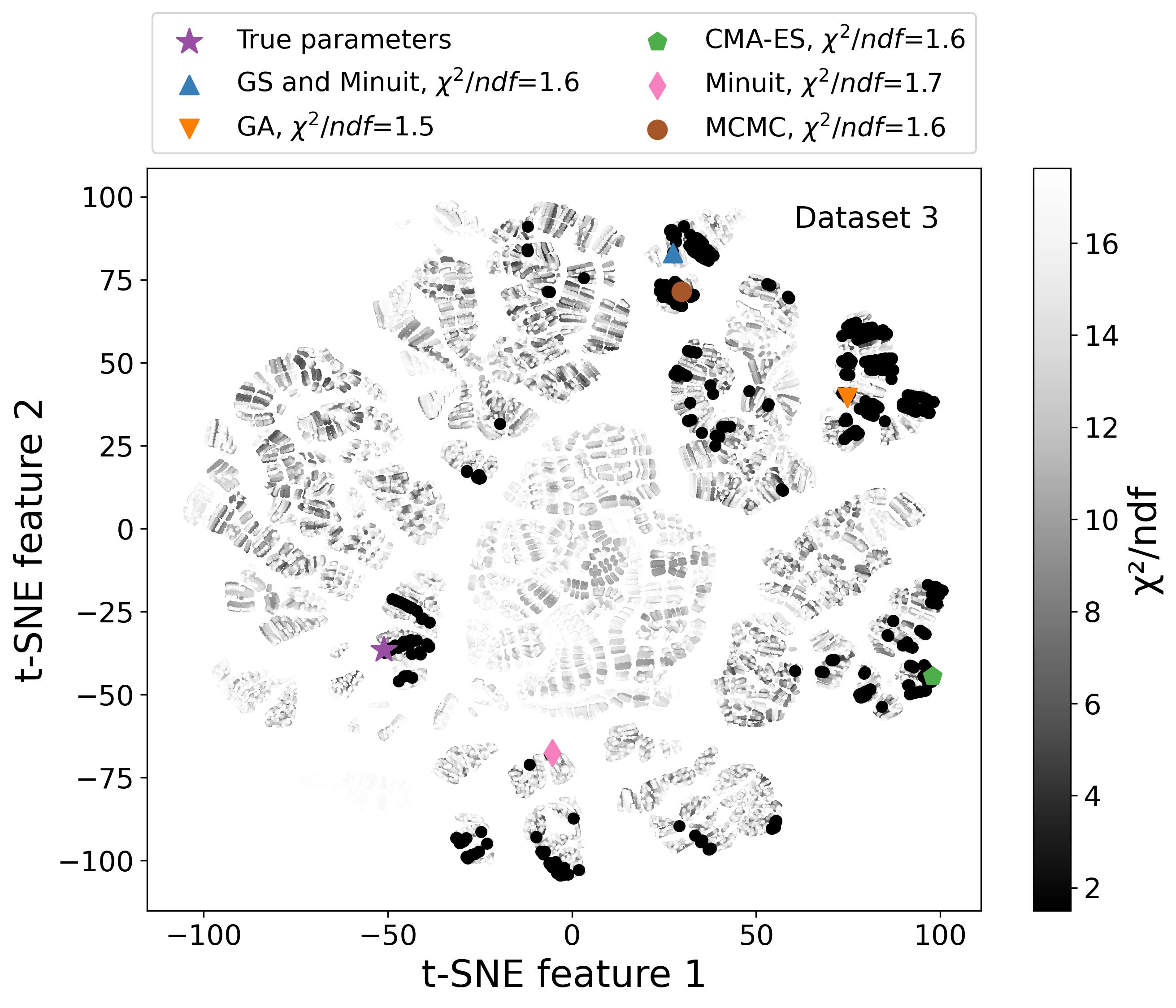}
         
         \label{fig:three sin x}
     \end{subfigure}
        \caption{Same as Fig. \ref{fig:results_simdata1} but for dataset 3.}
        \label{fig:results_simdata3}

\end{figure*}

The methods described above were applied to the simulated datasets described above to evaluate and compare their performance. Additionally, t-SNE was used to visualize the location of the found best-fit parameters from all optimization algorithms in the global parameter space. For this, their location, together with the location of the true parameters, was mapped together with the $50\,000$ lowest reduced $\chi^2$ points from the grid scan.  In the t-SNE plots, points that are assigned to $\chi^{2}/$ndf values in the interval of interest [($\chi^{2}/$ndf)$_\mathrm{{min}}$, ($\chi^{2}/$ndf)$_\mathrm{{min}}$ + 2] are highlighted as darker, larger points. The best-fit SEDs and their corresponding t-SNE maps are shown in Fig. \ref{fig:results_simdata1} -- \ref{fig:results_simdata3}. The best-fit model parameters found by each optimization method are listed in Table \ref{table:table_simdata}. In the first row of Table \ref{table:table_simdata}, the true parameters are presented for comparison with the results obtained from the five algorithms.

Fig. \ref{fig:results_simdata1} shows the results of the SED fitting for dataset 1. As shown in the left panel of Fig. \ref{fig:results_simdata1}, all the selected minimization algorithms found solutions that, overall, describe the data well. The solutions found with CMA-ES and the genetic algorithm have reduced $\chi^2$ values of 1.6 and 1.8, respectively. As can be seen in Table \ref{table:table_simdata}, their best-fit parameters agree with the true values within the parameter uncertainty ranges. The grid scan, MCMC, and  Minuit provide solutions with $\chi^2$/n.d.f. of 2.2, 3.5, and 4.1, respectively. As the lower part of the left panel in Fig. \ref{fig:results_simdata1} shows, they deviate from the true model mostly in the X-ray -- MeV region and in the radio band. The right panel of Fig. \ref{fig:results_simdata1} shows the location of the found best-fit model parameters in the global parameter space. As expected, the solutions from the grid scan, the genetic algorithm, and CMA-ES lie close to the true parameters. The other results from the grid scan (before the local optimization) with low values of reduced $\chi^2$ (marked as larger black dots) highlight only one region. Both the MCMC and Minuit solutions lie outside of this region. This performance issue could be caused by our setup and would likely lead to the same close results provided by the algorithms if the algorithms were run from another initial point or with a smaller step size. Overall, in this overly optimistic case of an almost perfect SED with small uncertainties of the photon flux values, the different minimization algorithms led to similar results (see Table~\ref{table:table_simdata}).

As shown with the grid scan, the only region in the parameter space with low $\chi^2$/n.d.f. was the one containing the true solution.

Similarly, Fig. \ref{fig:results_simdata2} shows the results of the SED fitting for dataset 2. Despite the overall good characterization of the data points with the best-fit model, the random deviations in the IR and optical ranges made the location of the synchrotron peak ambiguous. As shown in the left panel of Fig. \ref{fig:results_simdata2}, all found best-fit solutions predict different photon fluxes in the low-energy range. Additionally, the solutions from the grid scan ($\chi^2$/n.d.f. = 1.6) and CMA-ES ($\chi^2$/n.d.f. = 1.0) deviate from the Minuit ($\chi^2$/n.d.f. = 1.4), MCMC ($\chi^2$/n.d.f. = 1.3) and genetic algorithm ($\chi^2$/n.d.f. = 1.1) solutions in the keV - MeV range where the data coverage is poor. As shown in the right panel of Fig. \ref{fig:results_simdata2}, the closest solutions to the true model were found by MCMC, Minuit, and the genetic algorithm.  The values of the model parameters in the best-fit solution from the genetic algorithm and MCMC are also the closest to the true values when comparing their values with the obtained uncertainties to the true values. Both the CMA-ES and grid scan result in lower values of the electron luminosity compared to the true solution, which is compensated by a higher Lorentz factor and varying values of the blob radius and the magnetic field strength. Interestingly, CMA-ES, which yielded the best goodness of fit among all algorithms, is located the furthest from the true model, with none of the parameters agreeing with the true values within their uncertainties.  Apart from three regions in the global parameter space where the algorithms have found their best-fit solutions, the grid scan suggests additional regions with decent solutions marked by the large black points in the right panel of Fig. \ref{fig:results_simdata2}. 

The results of fitting dataset 3 are presented in Fig. \ref{fig:results_simdata3}. As expected, with fewer data points and a worse characterization of the SED features, the degeneracy of the model parameters increases. The best-fit models, shown in the left panel of Fig. \ref{fig:results_simdata3}, all yielded almost the same values of reduced $\chi^2$, yet the shapes of the best-fit SEDs vary greatly. Similar to dataset 2, the biggest difference is observed in the low-energy ($< 0.1$ eV) and MeV ranges. Additionally, due to the larger uncertainties and resulting larger scattering of the data points in GeV gamma rays, the characterization of the high-energy peak becomes challenging for this dataset. As shown in the right panel of Fig. \ref{fig:results_simdata3}, the best-fit solutions are scattered across the parameter space, and none of them is close to the true model. The grid scan results suggest that apart from the found solutions there exist multiple other regions in the parameter space that could produce similarly good solutions. Apart from the worsened characterization of the best-fit location in the global parameter space, larger flux uncertainties and fewer data have led to a worse determination of the minimum. As shown in Table \ref{table:table_simdata}, the uncertainties for the model parameters in dataset 3 are, on average, larger than for the previous datasets for all algorithms.

As shown with dataset 2 and dataset 3, the realistic SEDs are explained with degenerate models that result in a similar goodness of fit. With larger flux uncertainties and fewer data points, the degeneracy drastically increases. 
The models with the lowest $\chi^2$/n.d.f. values are not necessarily the closest to the true parameters. Only in the case of an almost perfect SED, the choice of the optimization algorithm does not affect the outcome.

\begin{table*}

\caption{Best-fit model parameters for simulated data.}             
\label{table:table_simdata}      

\centering                         

\begin{adjustbox}{width=\textwidth}
\begin{tabular}{c c c c c c c c c c}        

\toprule

Dataset & Method & lg($R_\mathrm{blob}^\prime$) [cm] & $B^\prime$ [G] & lg($\gamma_\mathrm{min}^\prime$) & lg($\gamma_\mathrm{max}^\prime$) & lg($L^\prime_\mathrm{e}~[erg~s^{-1}]$) & $\Gamma_\mathrm{b}$ & index $\alpha$ & $\chi^{2}/$ndf \\    
\midrule
True& &  16.58 & 0.41 & 3.30  & 4.88 & 43.39 & 7.8 & 2.82& - \\
\midrule                      

Dataset 1 &  Grid scan  & 16.477$\pm$0.001 & 0.42$\pm$0.03 & 3.234$\pm$0.001  & 4.98$\pm$0.03 & 43.202$\pm$0.004 & 9.2$\pm$0.1 & 2.88$\pm$0.06 & 2.2 \\      

  & GA & $16.60^{+0.07}_{-0.06}$  & 0.41$\pm$0.08 & 3.30$\pm$0.03 & 4.98$\pm$0.04 & $43.39^{+0.14}_{-0.22}$ & 7.8$\pm$0.5& 2.85$\pm$0.07 & 1.8\\
  & CMA-ES & 16.59$\pm$0.01 & 0.39$\pm$0.03 & 3.34$\pm$ 0.01 & 5.00$\pm$0.01& 43.365$\pm$0.003& 8.00$\pm$0.02& 2.890$\pm$0.003& 1.6 \\
  & Minuit & 16.13$\pm$0.05 & 0.51$\pm$0.08 & 3.05$\pm$0.53 & 4.71$\pm$0.10 & 42.79$\pm$0.04 & 12.3$\pm$2.83 & 2.80$\pm$0.04 & 4.1  \\
  & MCMC & $16.29^{+0.07}_{-0.08}$ & $0.36^{+0.07}_{-0.06}$ & $3.04^{+0.04}_{-0.03}$ &  $4.67\pm 0.03$ &  $42.84\pm0.04$ & $12.37^{+0.13}_{-0.17}$ & $2.77\pm0.06$ & 3.5 \\
\midrule
Dataset 2 & Grid scan  & 16.50$\pm$0.67 & 0.24$\pm$0.09 & 3.07$\pm$0.04 & 4.88$\pm$0.50 & 43.00$\pm$0.70 & 11.5$\pm$0.80 & 2.82$\pm$0.45 & 1.6 \\
 &GA & $16.34^{+0.10}_{-0.13}$& 0.54$\pm$0.12 & $3.28^{+0.07}_{-0.08}$ & $4.83^{+0.11}_{-0.15}$& $43.20^{+0.15}_{-0.22}$& 8.6$\pm$0.85 & 2.82$\pm$0.13 & 1.1\\
 & CMA-ES & 16.11$\pm$0.01& 0.45$\pm$0.01 & 3.00$\pm$0.01 & 4.66$\pm$0.01 & 42.79$\pm$0.01 &  12.40$\pm$0.06& 2.74$\pm$0.01& 1.0 \\
  & Minuit &16.310$\pm$0.003 & 0.62$\pm$0.02&3.3016$\pm$0.0004 &4.84$\pm$0.02 & 43.21$\pm$0.01&8.474$\pm$0.001 & 2.89$\pm$0.01&  1.4\\
  & MCMC & $16.41^{+0.12}_{-0.14}$ & $0.51^{+0.17}_{-0.08}$ & $3.31^{+0.03}_{-0.05}$ & $4.83^{+0.08}_{-0.09}$ & $43.26^{+0.09}_{-0.06}$ & $8.21^{+0.29}_{-0.32}$ & $2.84^{+0.10}_{-0.13}$ & 1.3\\
  \midrule
Dataset 3 & Grid scan  &15.88$\pm$0.06 &1.20$\pm$0.01 &3.21$\pm$0.02 & 4.39$\pm$0.31 & 43.06$\pm$0.11 & 9.0$\pm$0.17 & 2.83$\pm$0.25 & 1.6 \\
 &GA & 15.47$\pm$1.15 & 1.32$\pm$0.33& $3.04^{+0.20}_{-0.38}$ & $4.34^{+0.25}_{-0.63}$& 42.63$\pm$0.92 & 12.9$\pm$2.3 & 2.91$\pm$0.09& 1.5\\
 & CMA-ES & 15.000$\pm$0.006 & 1.74$\pm$0.08 & 3.000$\pm$0.006 & 4.99$\pm$0.02 & 42.14$\pm$0.04 & 17.8$\pm$0.4 & 3.34$\pm$0.15 & 1.6 \\
  & Minuit &16.81$\pm$0.13 &0.14$\pm$0.02 & 3.58$\pm$0.26&5.50$\pm$0.80 &42.59$\pm$0.33 &11.3$\pm$0.44 &3.25$\pm$0.07 & 1.7 \\
  & MCMC & $16.21^{+0.22}_{-0.20}$ & $0.84^{+0.23}_{-0.21}$ & $3.29^{+0.09}_{-0.10}$ & $4.55^{+0.25}_{-0.27}$ & $43.23^{+0.15}_{-0.17}$ & $8.05^{+0.92}_{-1.07}$ & $2.84^{+0.18}_{-0.30}$ &  1.6 \\
  \bottomrule
\end{tabular}
\end{adjustbox}
\tablefoot{Explanation for parameter notation is given in Table \ref{tab:lep_pars}.}
\end{table*}

\section{Application to observational data} \label{sec:real_data}

\subsection{Selected sources}
As a next step, we studied the performance of the same selected fitting procedures when applied to observational data of two blazars, PKS~0735+178 and Mrk 501. We selected those two sources, because the SED of PKS~0735+178 represents a frequent case of a source with limited available data, while the SED of Mrk 501, one of the most monitored blazars, represents a case of excellent data coverage. 

PKS 0735+178 is an intermediate-frequency-synchrotron-peaked (ISP) BL Lac object with an estimated redshift of z=0.45 \citep{2012A&A...547A...1N}. For this work, we used the gamma-ray quiescent state data from \cite{2024arXiv240904165O}. All the observational data were measured between January 23 and February 2, 2010. Since this source and its leptonic model from \cite{2024arXiv240904165O} was a prototype for our simulated data, the boundaries for the parameter space remain the same as the ones presented in Table \ref{tab:lep_pars_bound}.

Mrk 501 is a well-known high-frequency synchrotron-peaked (HSP) BL Lac object located at the redshift z = 0.034 \citep{1975ApJ...198..261U}. For the modeling, we considered the data taken during the quiescent period of Mrk 501 in 2017 -- 2020 from \cite{Abe_2023}. As argued in \cite{Abe_2023}, the measurements in radio and optical indicate low variability in these frequencies and are interpreted to originate in the outer regions of the jet (as opposed to the compact region we intend to model). We, therefore, treat these points as upper limits. 
Since Mrk 501 is a HSP source, its model parameters are expected to lie in a different region of the parameter space compared to the ISP source PKS 0735+178. 
As a reference for the typical values of one-zone leptonic model parameters of this source, we used the modeling results of \cite{Abe_2023} and limited our parameter space according to them. The defined parameter space boundaries for Mrk 501 are shown in Table \ref{tab:lep_pars_grid_scan_mrk501}. For consistency with \cite{Abe_2023}, the EBL model based on \cite{2008A&A...487..837F} is assumed.

\begin{table*}
\caption{Parameters resulting from different parameter search methods for the data of PKS 0735+178 from 2010.}             

\label{table:pksparams}      

\centering                          

\begin{tabular}{c c c c c c c c c}        
\toprule               

Method & lg($R_\mathrm{blob}^\prime$) [cm] & $B^\prime$ [G] & lg($\gamma_\mathrm{min}^\prime$) & lg($\gamma_\mathrm{max}^\prime$) & lg($L^\prime_\mathrm{e}~[erg~s^{-1}]$) & $\Gamma_\mathrm{b}$ & index $\alpha$ & $\chi^{2}/$ndf \\      
\midrule

 Grid scan & 15.27$\pm$0.04 & 1.26$\pm$0.37 & 3.221$\pm$0.005  & 5.17$\pm$0.27 & 42.47$\pm$0.42 & 14.9$\pm$2.48 & 3.27$\pm$0.54 & 0.9 \\      
   
   GA & $15.54^{+0.06}_{-0.07}$ & 1.54$\pm$0.10 & $3.41^{+0.04}_{-0.05}$ & $4.96^{+0.05}_{-0.06}$ & $42.96^{+0.16}_{-0.27}$  & 9.65$\pm$0.60 & 3.30$\pm$0.08 & 1.0\\
   CMA-ES & $15.18^{+0.03}_{-0.04}$ & 1.24$\pm$0.29& 3.00$\pm$0.02 & $4.85^{+0.08}_{-0.10}$ & $42.21^{+0.03}_{-0.04}$ & 18.2$\pm$0.46 & 3.17$\pm$0.07 & 1.1 \\
   
   Minuit & 15.082$\pm$0.004 & 1.71$\pm$0.01 & 3.09$\pm$0.03 & 5.15$\pm$0.36 & 42.23$\pm$0.03 & 16.6$\pm$0.26 & 3.26$\pm$0.06 & 1.3  \\
   MCMC & $15.71^{+0.16}_{-0.12}$ & $1.22\pm0.14$ & $3.36\pm0.05$ & $4.84^{+0.18}_{-0.15}$ & $42.92\pm 0.07$ & $9.65^{+0.47}_{-0.35}$ & $3.08^{+0.16}_{-0.13}$ & 2.3 \\
\bottomrule

\end{tabular}
\tablefoot{Explanation for parameter notation is given in Table \ref{tab:lep_pars}.}
\end{table*}

\begin{table*}
\caption{Parameters resulting from different parameter search methods for the data of Mrk 501.}             
\label{table:params_mrk501}      

\centering 
\begin{adjustbox}{width=\textwidth}

\begin{tabular}{c c c c c c c c c}       

\toprule                 

Method & lg($R_\mathrm{blob}^\prime$) [cm] & $B^\prime$ [G] & lg($\gamma_\mathrm{min}^\prime$) & lg($\gamma_\mathrm{max}^\prime$) & lg($L^\prime_\mathrm{e}~[erg~s^{-1}]$) & $\Gamma_\mathrm{b}$ & index $\alpha$ & $\chi^{2}/$ndf \\   
\midrule

 Grid scan & 15.522$\pm$0.007 & 0.32$\pm$0.03 & 3.22$\pm$0.34 & 5.256$\pm$0.0004 & 40.63$\pm$0.02 & 12.5$\pm$0.02 & 1.97$\pm$0.05 & 29.7 \\      
   
   GA & $15.61^{+0.07}_{-0.08}$ & 0.22$\pm$0.03 & $3.14^{+0.11}_{-0.15}$ & $5.38^{+0.04}_{-0.05}$ & $40.65^{+0.10}_{-0.12}$ & 15.1$\pm$0.56 & 2.26$\pm$0.06 & 28.9 \\
   CMA-ES & $16.08^{+0.03}_{-0.04}$ & 0.15$\pm$ 0.01 & $2.79^{+0.15}_{-0.22}$ & 5.51$\pm$ 0.02 & 41.13$\pm$ 0.04 & 10.2$\pm$ 0.3 & $2.04^{+0.03}_{-0.12}$ & 28.9 \\
   
   Minuit & 15.485$\pm$0.005 & 0.286$\pm$0.004 & 2.378$\pm$0.005 & 5.231$\pm$0.005 & 40.699$\pm$0.003 & 13.44$\pm$0.05 & 1.959$\pm$0.002 &  33.9 \\
   MCMC & $15.63\pm 0.05$ &  $0.22^{+0.03}_{-0.02}$&  $1.97^{+0.21}_{-0.13}$ & $5.33\pm 0.03$ & $40.91^{+0.05}_{0.06}$ & $13.21^{+0.20}_{-0.14}$ & $2.05^{+0.03}_{-0.02}$ & 27.7 \\
   
\bottomrule

\end{tabular}
\end{adjustbox}
\tablefoot{Explanation for parameter notation is given in Table \ref{tab:lep_pars}.}
\end{table*}

\subsection{Results}
\subsubsection{PKS 0735+178}

\begin{figure}

\centering
     \begin{subfigure}{0.45\textwidth}
         \centering
         \includegraphics[width=\textwidth]{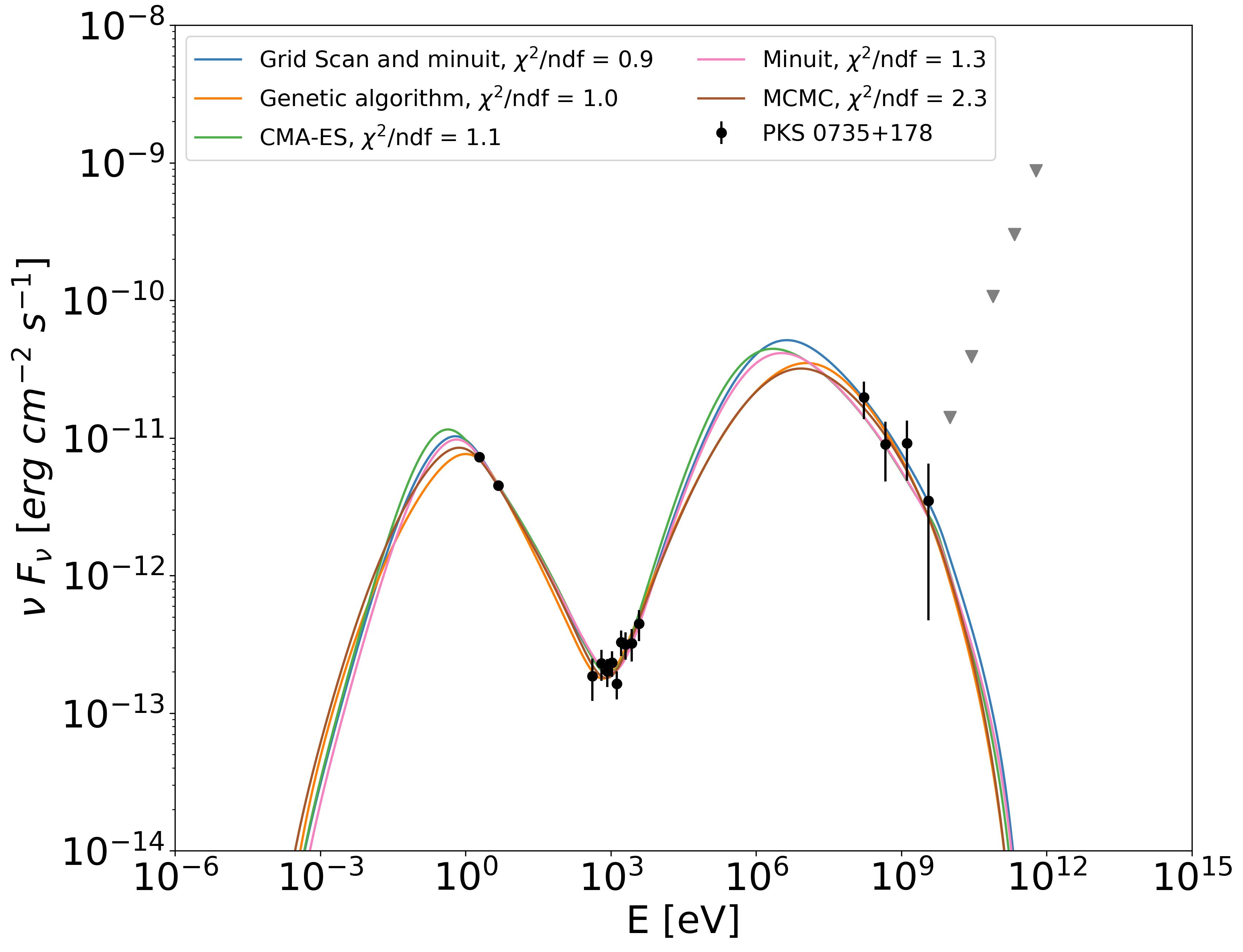}
         
         \label{fig:pks_models}
     \end{subfigure}
     \begin{subfigure}{0.45\textwidth}
         \centering
         \includegraphics[width=\textwidth]{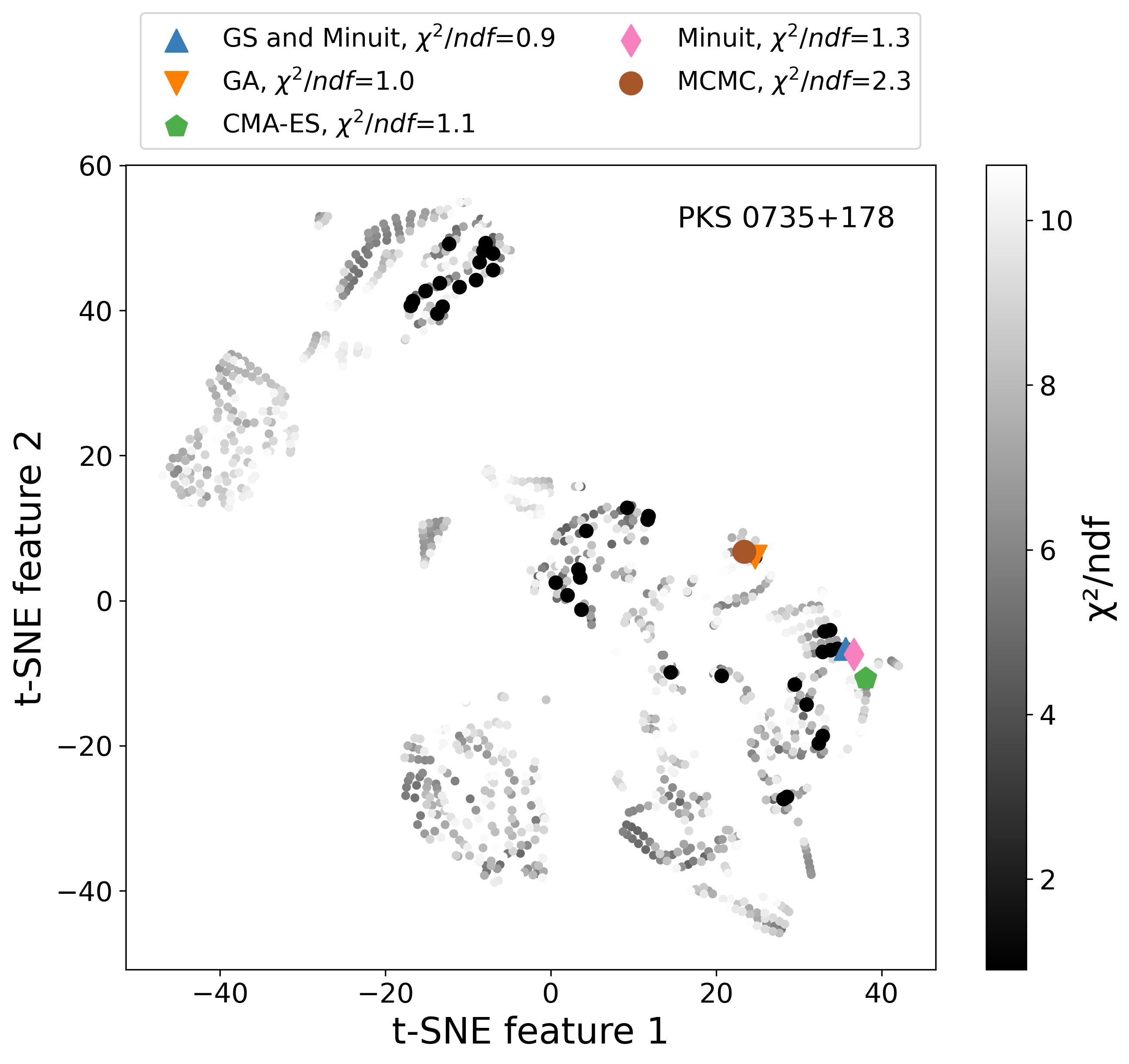}
         
         \label{fig:pks_tsne}
     \end{subfigure}
     
        \caption{Results of the SED fitting for PKS 0735+178. Upper panel: the best-fit results from all selected optimization algorithms. Lower panel: the location of best-fit solutions in the global parameter space shown as t-SNE map.}
        \label{fig:pks_results}

\end{figure}

The results of the SED fitting for PKS 0735+178 along with the corresponding t-SNE map of the best-fit parameter location are shown in Fig.~\ref{fig:pks_results}. The values of the best-fit model parameters for each minimization procedure are listed in Table \ref{table:pksparams}. With the exception of the solution found by the MCMC algorithm, all methods yielded fits with $\chi^2$/n.d.f. values close to one, indicating that they provide equally good explanations for the dataset. MCMC, by contrast, resulted in a solution with $\chi^2$/n.d.f.=2.3. The higher value is primarily driven by the second data point, which has a small uncertainty. Even small deviations from this point can lead to a large contribution to the $\chi^2$/n.d.f. value. Regardless of the goodness-of-fit values, we observe in Fig.~\ref{fig:pks_results} that the resulting curves differ in regions where no data are available.

As shown in the lower panel of Fig. \ref{fig:pks_results}, the found model parameters lie in different regions of the parameter space. While the best results from the grid scan ($\chi^2$/n.d.f. = 0.9), Minuit ($\chi^2$/n.d.f. = 1.3) and CMA-ES ($\chi^2$/n.d.f. = 1.1) are closer to each other, the MCMC ($\chi^2$/n.d.f. = 2.3) and the genetic algorithm ($\chi^2$/n.d.f. = 1.0) results are outside of this cluster. This can also be seen in the upper panel of Fig. \ref{fig:pks_results}, where we observe that the solutions provided by MCMC and the genetic algorithm deviate from the other three models, particularly in the radio and MeV domains, where no data are available to constrain the fits.

The t-SNE map shows that the grid scan, the CMA-ES and the Minuit results lie in those clusters where also the grid scan finds regions of favorable solutions. Interestingly, apart from the found solutions, the grid scan also suggests multiple regions of parameter space that could yield good solutions highlighting how degenerate the parameter space is for this SED.

The SED data of PKS 0735+178 during its quiescent state is a good example of a degenerate case since the features of the SED, such as the location of the synchrotron or the high-energy peaks as well as the peak flux values, cannot be constrained. This, in turn, leads to great uncertainties in most of the model parameters. Similar to simulated dataset 3, almost all solutions have close $\chi^{2}$/n.d.f. values despite their different best-fit model parameter values. The genetic algorithm and MCMC results deviate strongly from the other solutions. They suggest a scenario with a larger blob and higher electron luminosity. Moreover, while the other models infer bulk Lorentz factors up to $\Gamma_{b} > 14$, the genetic algorithm and MCMC both yield the same lower value of $\Gamma_{b} = 9.65$.

\subsection{Mrk 501}

\begin{figure}
\centering
     
     \begin{subfigure}{0.45\textwidth}
         \centering
         \includegraphics[width=\textwidth]{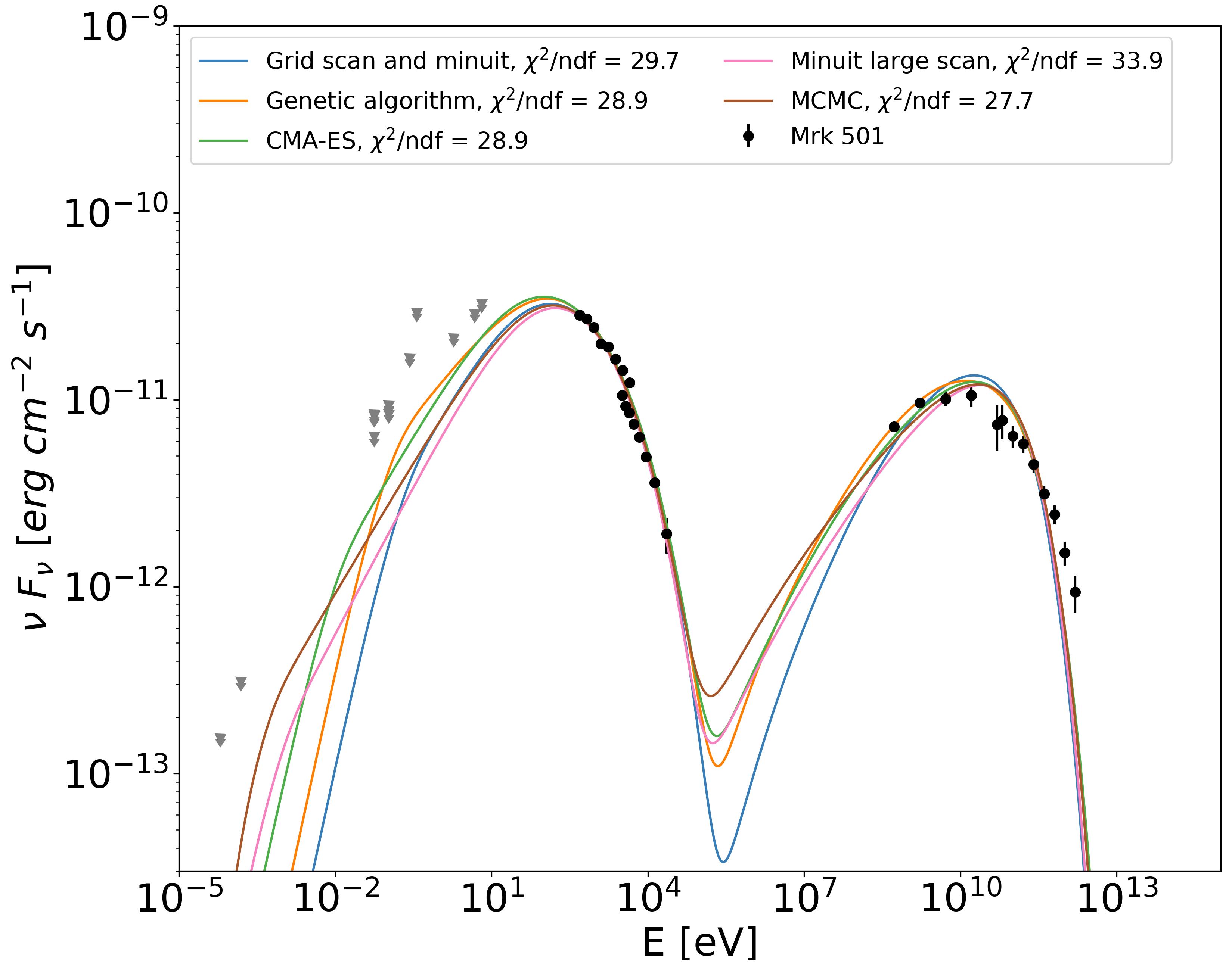}
         
         \label{fig:mrk_tsne}
     \end{subfigure}
     \begin{subfigure}{0.45\textwidth}
         \centering
         \includegraphics[width=\textwidth]{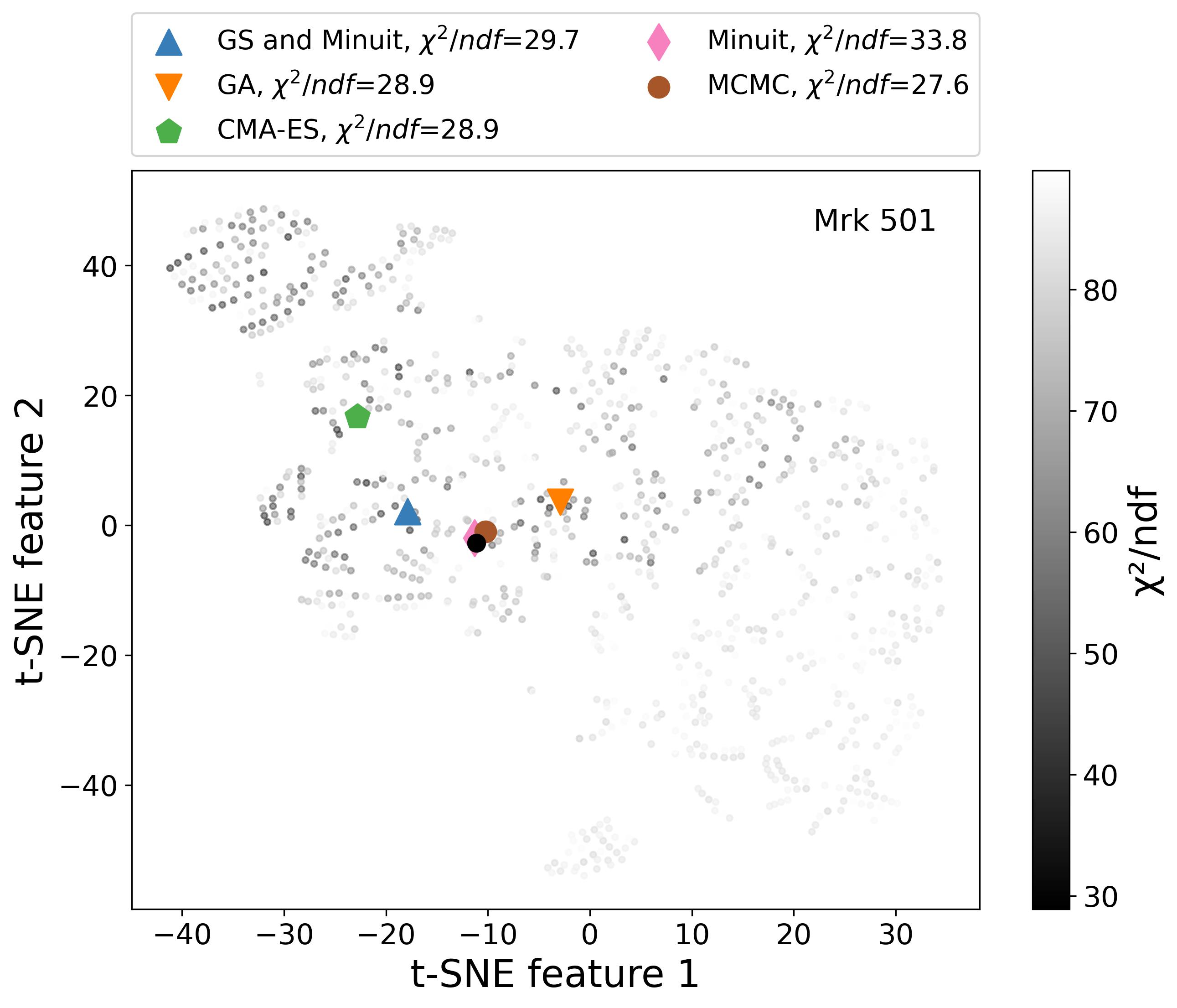}
         
         \label{fig:mrk_models}
     \end{subfigure}
     
        \caption{Results of the SED fitting for Mrk 501.  Upper panel: the best-fit results from all selected optimization algorithms. Data used for the modeling is represented as black points while the grey triangles are treated as upper limits. The discontinuity visible within the X-ray energy range could be due to the fact that these data were measured by two different instruments. Lower panel: the location of the best-fit solutions in the global parameter space shown as t-SNE map.}
        \label{fig:mrk_results}

\end{figure}

The results of the SED fitting for Mrk 501 are shown in the upper panel of Fig. \ref{fig:mrk_results}, while the lower panel of Fig. \ref{fig:mrk_results} shows the t-SNE map with the best-fit locations in the parameter space. The SED of Mrk 501 built for the quiescent state has more data points than the SED of PKS 0725+178. Still, interpreting optical and radio data for Mrk 501 as upper limits makes it challenging to constrain the location of the synchrotron peak. In addition, the gap between the low-energy and high-energy peaks falls in the MeV range with no available data. 

All of the found solutions fit the X-ray data well. While they also match the high-energy flux levels, this type of model fails to correctly reproduce the shape of the gamma-ray spectrum which also leads to high absolute values of the reduced $\chi^2$ ($>26$). This may be an indication that more complex model assumptions are required for this source. We assumed a simple power-law energy spectrum of the pre-accelerated electrons. As electrons cool due to synchrotron emission, the cooling break appears self-consistently in our modeling. However, \cite{Abdo_2011} were able to successfully reproduce the Mrk 501 SED only with the introduction of multiple breaks in the electron spectrum. Alternatively, the parameters that could produce a better fit could lie outside of the considered region in the parameter space (see Table \ref{tab:lep_pars_grid_scan_mrk501}).

Even though the obtained values of the reduced $\chi^2$ cannot be considered as a good fit, the upper panel of Fig.~\ref{fig:mrk_results} shows how different minimization algorithms produced independent solutions. All the best-fit solutions predict a very similar shape of the X-ray fluxes but diverge in their predictions on the behavior in the dip between the two SED peaks.

The t-SNE plot (lower panel of Fig.~\ref{fig:mrk_results}) reveals that the best-fit parameters found by MCMC ($\chi^2$/n.d.f. = 27.7) and Minuit ($\chi^2$/n.d.f. = 33.9) are similar, while the solutions of CMA-ES and the grid scan belong to different families of solutions. Unlike the grid scan results from PKS 0735+178 with multiple regions of low $\chi^2$/n.d.f values, the grid scan results for Mrk 501 show one prospective solution in the interval of interest (large black marker in the lower panel of Fig.~\ref{fig:mrk_results}, overlapping with the Minuit and MCMC solutions). The corresponding best-fit SEDs, as shown in the upper panel of Fig. \ref{fig:mrk_results}, agree only in the X-ray and high-energy gamma-ray ranges and significantly deviate in their predictions in all other energy ranges. 

Despite the variations in parameters that could not be contained (minimal electron Lorentz factors, bulk Lorentz factor, and electron spectral index), most of the models require a blob radius of ${\sim}3\times 10^{15}$ cm, a magnetic field strength around 0.2--0.3 G, and an electron luminosity of ${\sim}5\times 10^{40}$ erg/s. The most discrepant solution was found by CMA-ES where the blob has a three times larger radius, the required magnetic field strength is lower by a factor of two, and the bulk Lorentz factor is lower, which is compensated by a higher electron luminosity. 

Overall, this case is in some respects similar to dataset 2, as there are relatively few models that provide a good fit to the data. The higher number of data points constrains the range of possible solutions, preventing the situation we encountered with dataset 3 or for PKS 0735+178, where, for instance, up to 20 distinct solutions were found within the parameter space of interest. However, despite this constraint, the parameter space remains degenerate. Even among the models that successfully reproduce the X-ray data, there are significant differences between the models, particularly in regions where observational data are absent.

\section{Discussion} \label{sec:discuss}
By modeling simulated and observed SEDs of blazars, we demonstrated the expected issue of parameter degeneracy in the simplest one-zone leptonic radiation models, which was described in Sec. \ref{sec:smoothness}.  This issue was mentioned in many previous works dedicated to blazar modeling. For instance, \cite{2017A&A...603A..31A} discussed the advantages of using a grid scan approach in the context of finding multiple solutions. 
 
\cite{lucchini2019breaking} addressed the parameter degeneracy of multi-zone models and proposed a way to break it by fitting six consecutive blazar states jointly. Many works, especially those utilizing MCMC, show the pairwise distributions of model parameters around the best-fit solutions. We note that the plots with such two-dimensional projections (using marginal distributions or some parameters fixed to their best-fit values) cannot capture any other significantly different solutions. In this work, we tackled this problem and showed how physically different solutions and their proximity in the parameter space can be compared using t-SNE. However, as already discussed in Section \ref{sec:tsne}, a careful choice of hyperparameters such as perplexity is required.

In Section \ref{sec:optimization}, we applied the grid scan, the genetic algorithm, CMA-ES, MCMC, and Minuit to simulated datasets to estimate how close the best-fit parameters found by each method are to the true parameters. Dataset 1 represented a case of perfect data with no gaps and small uncertainties. Both evolutionary algorithms accurately identified the best-fit parameters that were consistent with the true parameter values within the uncertainty interval. The grid scan followed by a local minimization delivered best-fit parameters a bit further from the true parameters compared to the evolutionary algorithms. Likely, in case of better grid discretization or local minimization within slightly broader boundaries and more function calls, the true parameters could have been reached. Both Minuit and MCMC stopped around the local minima. Given the complexity of finding only one solution in an extremely vast parameter space, these algorithms likely required more probed points in the parameter space than what was allocated by us. On the contrary, in more degenerative cases (dataset 2 and dataset 3), these methods were able to find good solutions due to the high probability of finding such a solution even when probing a small amount of points and covering a limited parameter space region. 
We note that nested sampling, such as the MultiNest algorithm used in \cite{2024ApJ...963...71B} and \cite{2024ApJ...971...70S}, may be a more efficient sampling approach.

In our setup, with ten probed points per parameter, the grid scan resulted in being the most computationally expensive method. However, it is the only known approach that enables the exploration of multiple physically different solutions. It also allowed us to visualize the parameter space with t-SNE due to comprehensive parameter space coverage. While our SSC one-zone model is the simplest model for explaining a blazar SED, requiring only seven free parameters, more complex models with a higher number of free parameters would require drastically higher computational resources. However, the possibility of reusing its results for other sources or datasets offers an advantage. A known limitation of the grid scan is that solutions may fall between predefined grid points. Some studies \citep[e.g.][]{2017A&A...603A..31A} address this by performing additional, finer grid scans on top of the initial large grid. While this approach can improve accuracy, it comes at the cost of increased computational time. In our approach, the second step was changed to a local minimization with Minuit.

The results showed how accurately the evolutionary algorithms captured the true solution. Both algorithms performed similarly, with CMA-ES converging much faster and offering the additional advantage of self-adapting its internal parameters. Both algorithms converge to only one solution. While it is possible to extract all solutions from each generation to potentially identify multiple solutions, this does not provide the same coverage as the grid scan. Evolutionary algorithms explore the parameter space based on evaluations, which can lead to certain regions being excluded from further consideration as the algorithms progressively focus on specific regions with each generation. This behavior is also reflected in the parameter uncertainties. We calculated the standard deviation of the parameters from the final generation, which yielded relatively small uncertainties, as the solutions in the final generation had already converged closely towards each other. We note, however, that the results of the minimization with the evolutionary algorithms are not guaranteed to be the same if the minimization is repeated with the same conditions. This is caused by the random initialization of the initial population and the highly irregular structure of the parameter space.

Apart from the case of a perfect dataset, our results from analyzing the parameter space of the leptonic radiation models suggest that, regardless of the method used, the complex nature of the parameter space with multiple optima can prevent any algorithm from capturing the true values. Especially for the models found with algorithms converging to a single solution, it can be challenging to accurately explain the emission of blazars. In the end, the best approach is to be aware of the existence of multiple solutions and to use methods like the grid scan (either a nested grid scan or combined with a local minimization), as it can represent the parameter space in the most comprehensive way rather than providing only a single solution. 

For the case of the observational data, we modeled the radiation of PKS~0735+178 during a quiescent state with each of the five minimization algorithms, leading to different solutions. This state was modeled before in \cite{2024MNRAS.529.3503B} and \cite{2024arXiv240904165O}. \cite{2024arXiv240904165O} addressed the issue of multiple possible solutions by performing a grid scan and selecting two physically different solutions (``slow'' and ``fast'' solutions based on the bulk Lorentz factor). Our best-fit models from all minimization algorithms have a two to three times higher magnetic field strength and a much lower emission zone radius compared to both the slow and fast models for this state in \cite{2024arXiv240904165O}. Similarly, in our solutions, the magnetic field strength is twice as high, and the blob radius is one order of magnitude lower than that in \cite{2024MNRAS.529.3503B}.  This indicates that all five best-fit solutions provide new models in addition to those found in the literature.

For Mrk 501, our results reproduce the overall flux levels but fail to match the gamma-ray spectral shape. The parameter boundaries were based on \cite{Abe_2023}, and the same dataset was used. While \cite{Abe_2023} fixed the blob size to $R'_{\textrm{blob}}$ = $10^{17.06}$ cm, our solutions suggest significantly smaller emission regions. The magnetic field strengths obtained from our searches were around $B' = 0.15-0.32$ G, whereas they suggest a field strength much lower with $B' = 0.025$ G. The bulk Lorentz factor was fixed to 11 in their work, which is comparable to our range of $\Gamma_b =10.2 - 15.1$. The required electron luminosity was also $10^3$ times lower in our case. Despite adopting parameter space boundaries from \cite{Abe_2023}, our results differ considerably. 
\cite{Abdo_2011} used a double broken power law to successfully reproduce the SED of Mrk 501 in the quiescent state which may be a reason why our simple power law (with naturally occurring cooling break) failed to reproduce the gamma-ray spectral shape.

Our findings, along with comparisons to other models, highlight the highly degenerate parameter space of the leptonic radiation models. Even the simplest SSC one-zone framework that has seven free parameters presents significant challenges in explaining the blazar emission. More complex models add additional parameters, further complicating the parameter space. This emphasizes the importance of careful interpretation of modeling results. To break this degeneracy, high-quality quasi-simultaneous multi-wavelength data are essential. In this sense, blazar surveys and monitoring programs are crucial to constrain the possible parameter values of the one-zone models. Additionally, exploring new energy ranges, such as MeV or TeV gamma rays, would significantly constrain the parameter space. For example, a single MeV data point added to our simulated dataset 3 would reduce the number of equally well-fitting models from five to one or two. Therefore, future MeV missions such as e\nobreakdash-ASTROGAM \citep{2017ExA....44...25D}, COSI \citep{2019BAAS...51g..98T} or AMEGO-X \citep{2022JATIS...8d4003C} would play an important role in breaking the model degeneracy. A similar effect is expected by adding data points in the TeV gamma-ray range, especially for HSP sources. The future  Cherenkov Telescope Array Observatory \cite[CTAO,][]{2019scta.book.....C} is expected to provide high-energy gamma-ray data, which will be essential for constraining the models. In the case of more complex leptohadronic models, multi-wavelength polarization can potentially constrain the hadronic component \citep{2024ApJ...967...93Z}.

\section{Summary and conclusions} \label{sec:concl}

In this work, we studied the fitting procedures of blazar SEDs in the context of one-zone leptonic models. We found convincing evidence that the goodness of fit is not a smooth and convex function of the model parameters due to the nature of the underlying radiation processes and parameter degeneracies.
Using the simulated pseudo-data, we observed that the degeneracies arise due to the missing data in certain energy ranges. An increase in the flux measurement uncertainty seems to further enhance these degeneracies.

We showed that for a typical blazar SED with data covering the optical, X-ray, and GeV gamma-ray ranges, the choice of the fitting procedure, in particular the choice of the minimizing algorithm, leads to considerably different results. In most cases, the five tested minimizing algorithms found physically different best-fit solutions. 

We applied the same fitting procedures to the observed SEDs of two blazars, PKS 0735+178 and Mrk 501. The model parameters could not be constrained for PKS 0735+178 due to unconstrained SED features such as the location of the synchrotron and the high-energy peak and the corresponding peak flux values. While the results for Mrk 501 were more consistent due to the better-characterized high-energy peak, the results of the different minimization algorithms were still degenerate, yet none of them yielded a satisfactory fit. 

The degeneracy is expected to become only worse when adding extra parameters to the models as in the case of, e.g., external radiation field models or when adding hadrons to the emission zone. This creates a challenge for gamma-ray and neutrino astronomy as retrieving source properties from the SED modeling becomes ambiguous. 

To reduce the parameter degeneracy it is crucial to assure a complete and simultaneous wavelength coverage of the data. Blazar monitoring programs as well as measurements in new energy ranges (e.g. MeV or very-high-energy gamma rays) could significantly improve the constraints on the model parameter space. Alternatively, taking into account multi-wavelength polarization provides another approach to break the degeneracy in lepto-hadronic models.

\begin{acknowledgements}
     F.A., A.F., and A.O. acknowledge the support from the DFG via the Collaborative Research Center SFB1491 \textit{Cosmic Interacting Matters - From Source to Signal}. A.O. was supported by DAAD funding program 57552340. F.A. acknowledges support by the Helmholtz Weizmann Research School on Multimessenger Astronomy. 
\end{acknowledgements}

\bibliographystyle{aa} 
\bibliography{biblio}

\begin{appendix} \label{sec:appendix}
\section{Perplexity and learning rate for t-SNE}  \label{sec:appendix:hyperparameters}
As mentioned in Section \ref{sec:tsne}, the hyperparameters for the application of t-SNE have to be selected carefully. The perplexity can be interpreted as the number of nearest neighbours for each point and has a significant impact on the result presentation. Another hyperparameter is the learning rate, which defines the step size for the gradient descent method in the Kullback-Leibler divergence minimization.

\begin{figure*}

\centering
     \begin{subfigure}{0.33\textwidth}
         \centering
         \includegraphics[width=\textwidth]{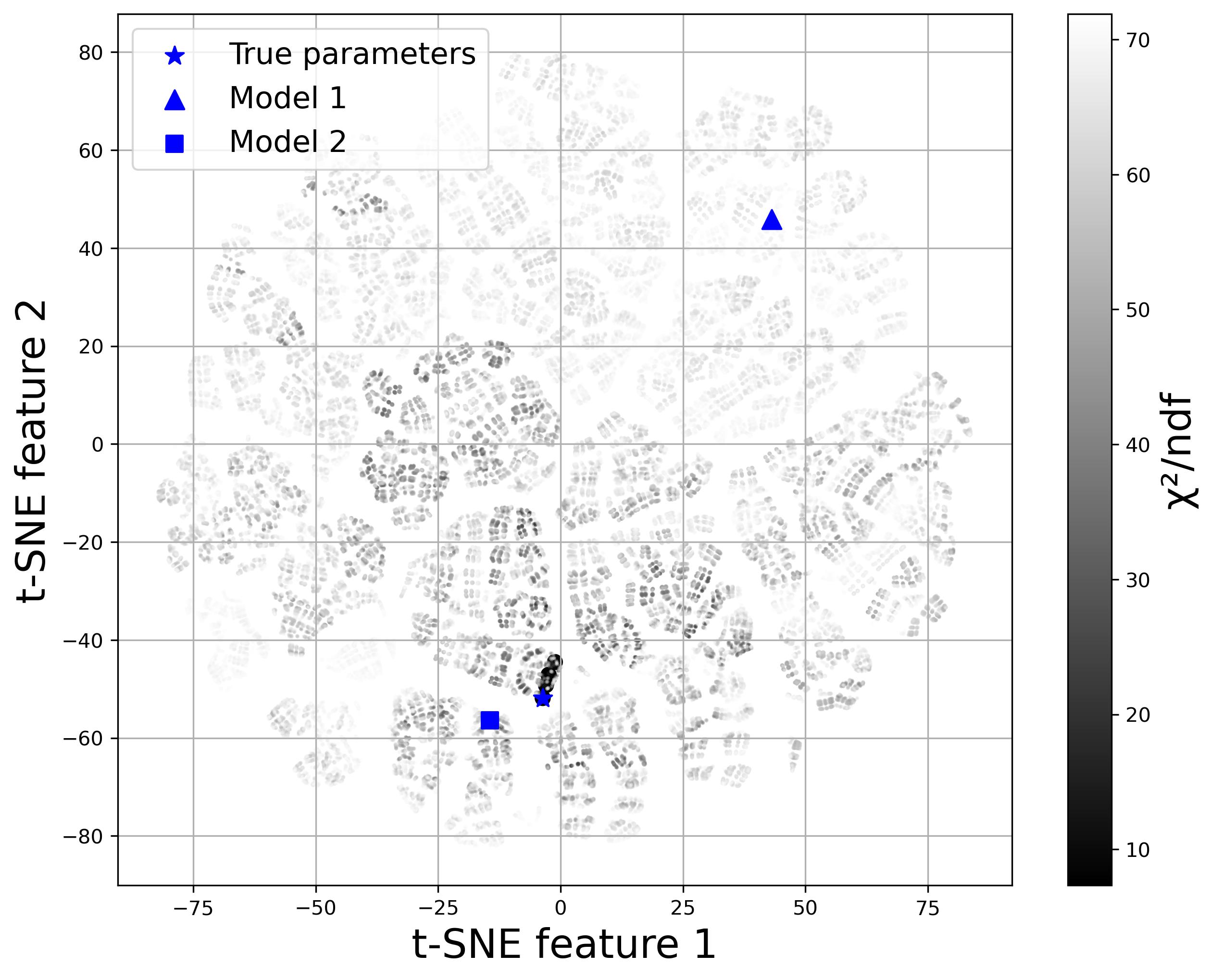}
         
         \label{fig:y equals x}
     \end{subfigure}
     \begin{subfigure}{0.33\textwidth}
         \centering
         \includegraphics[width=\textwidth]{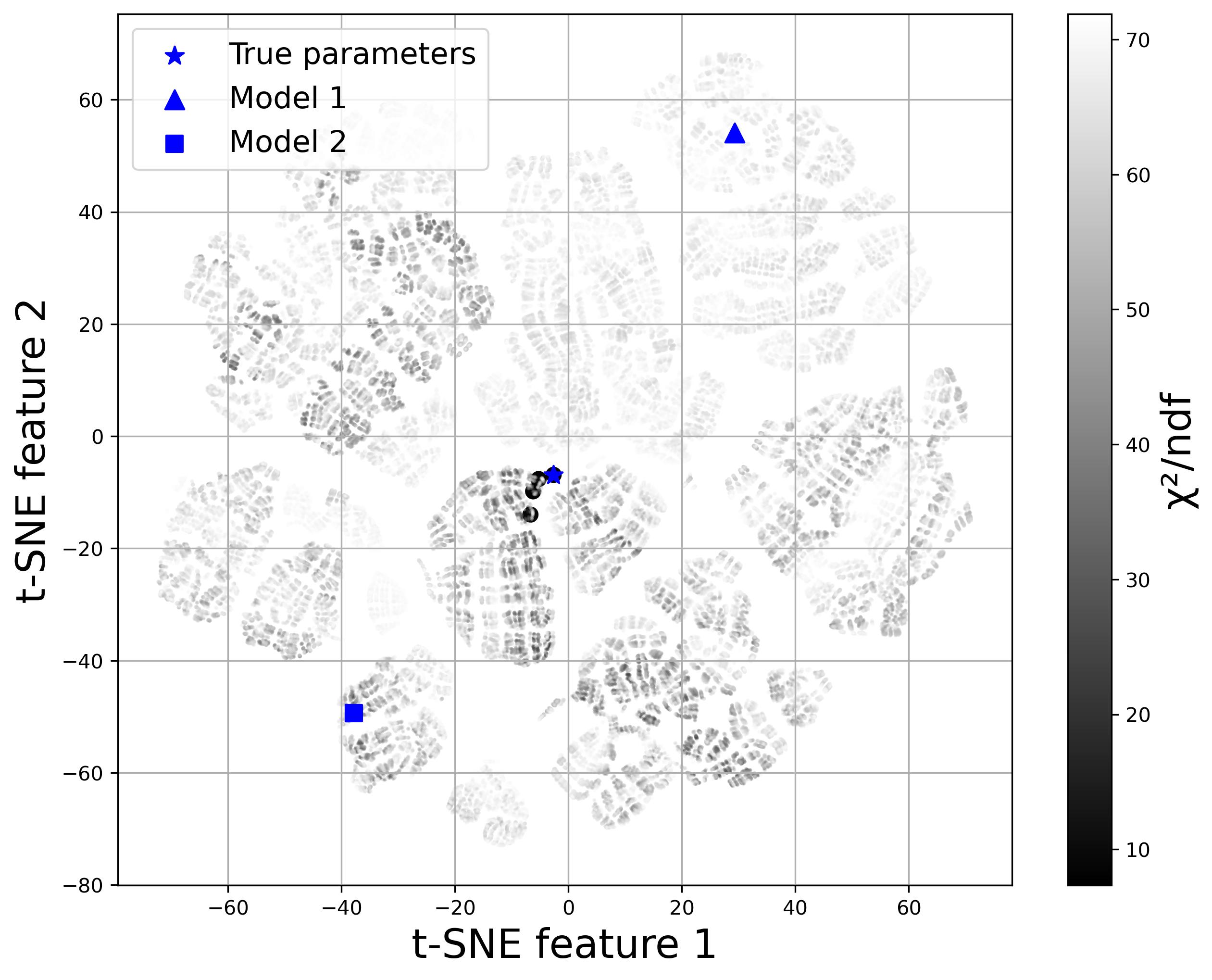}
         
         \label{fig:three sin x}
     \end{subfigure}
      \begin{subfigure}{0.33\textwidth}
         \centering
         \includegraphics[width=\textwidth]{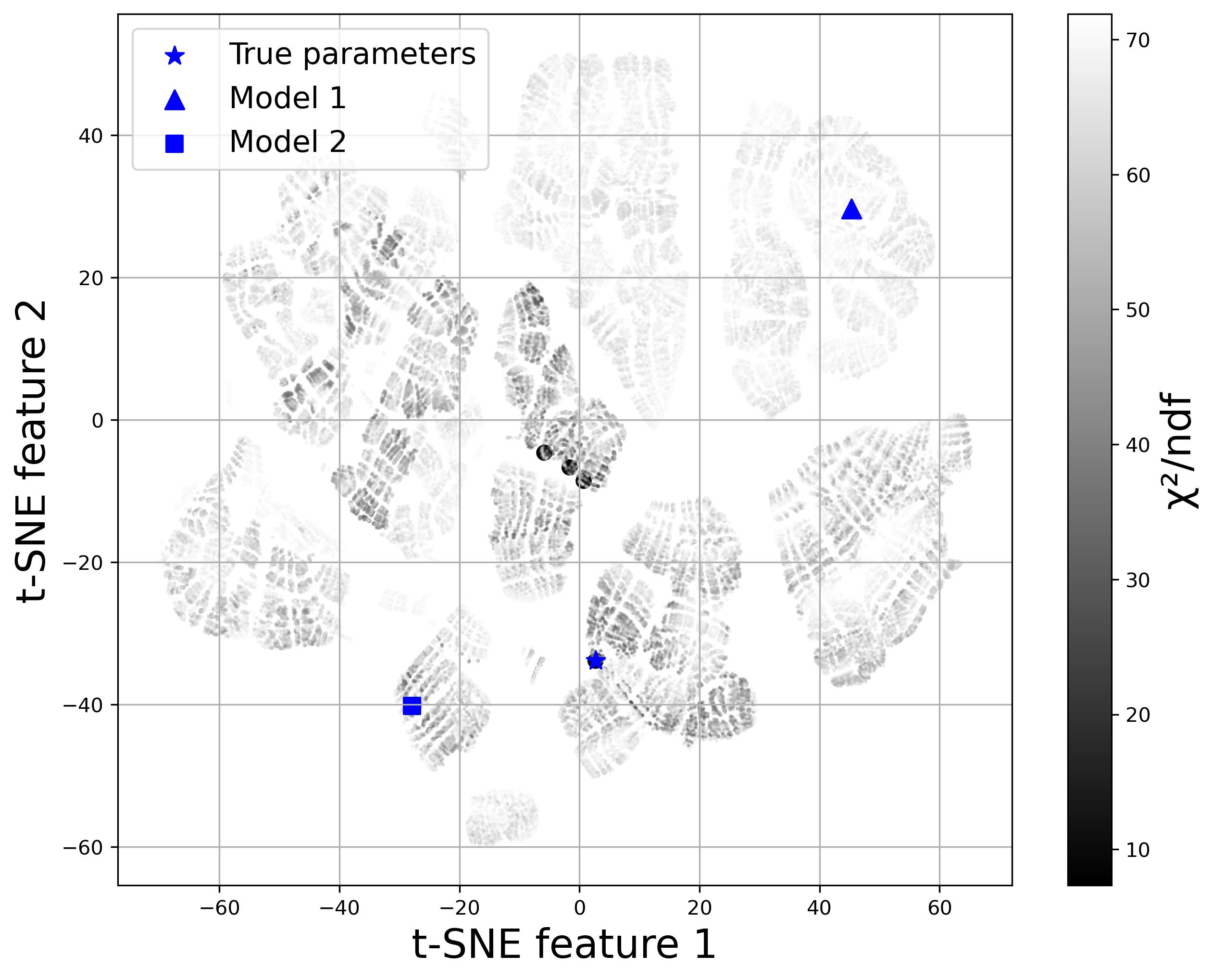}
        
         \label{fig:five over x}
     \end{subfigure}
        \caption{Three t-SNE plots with different perplexity values. Model 1 and Model 2 represent models that are significantly different from the true solution. The black dots are solutions with small $\chi^{2}/$ndf values that are close to the true parameters. The plots show the cases of perplexity values $p$=10, 30 and 80 in this order.}
        \label{fig:perplexity}

\end{figure*}

\begin{figure*}

\centering
     \begin{subfigure}{0.33\textwidth}
         \centering
         \includegraphics[width=\textwidth]{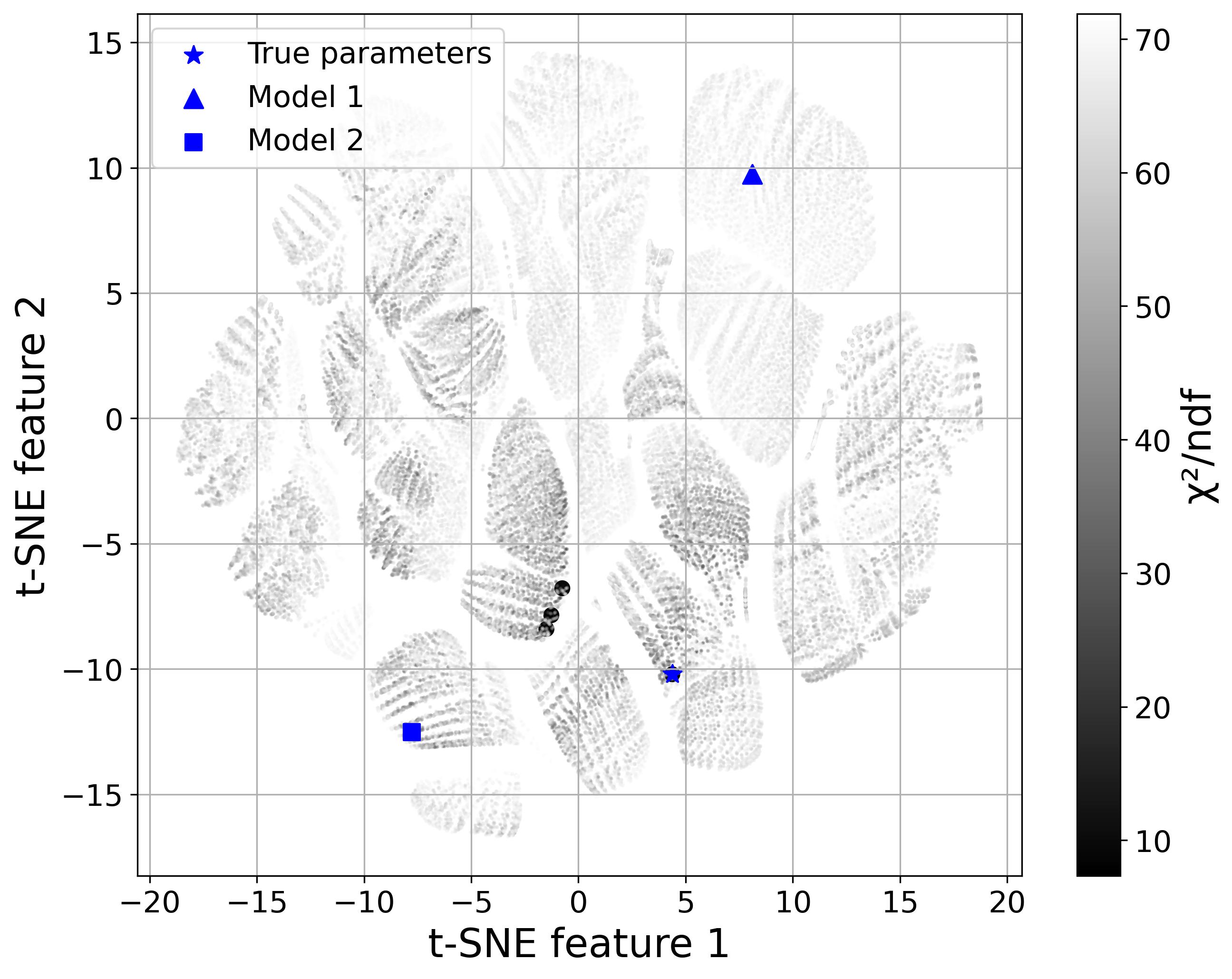}

     \end{subfigure}
     \begin{subfigure}{0.33\textwidth}
         \centering
         \includegraphics[width=\textwidth]{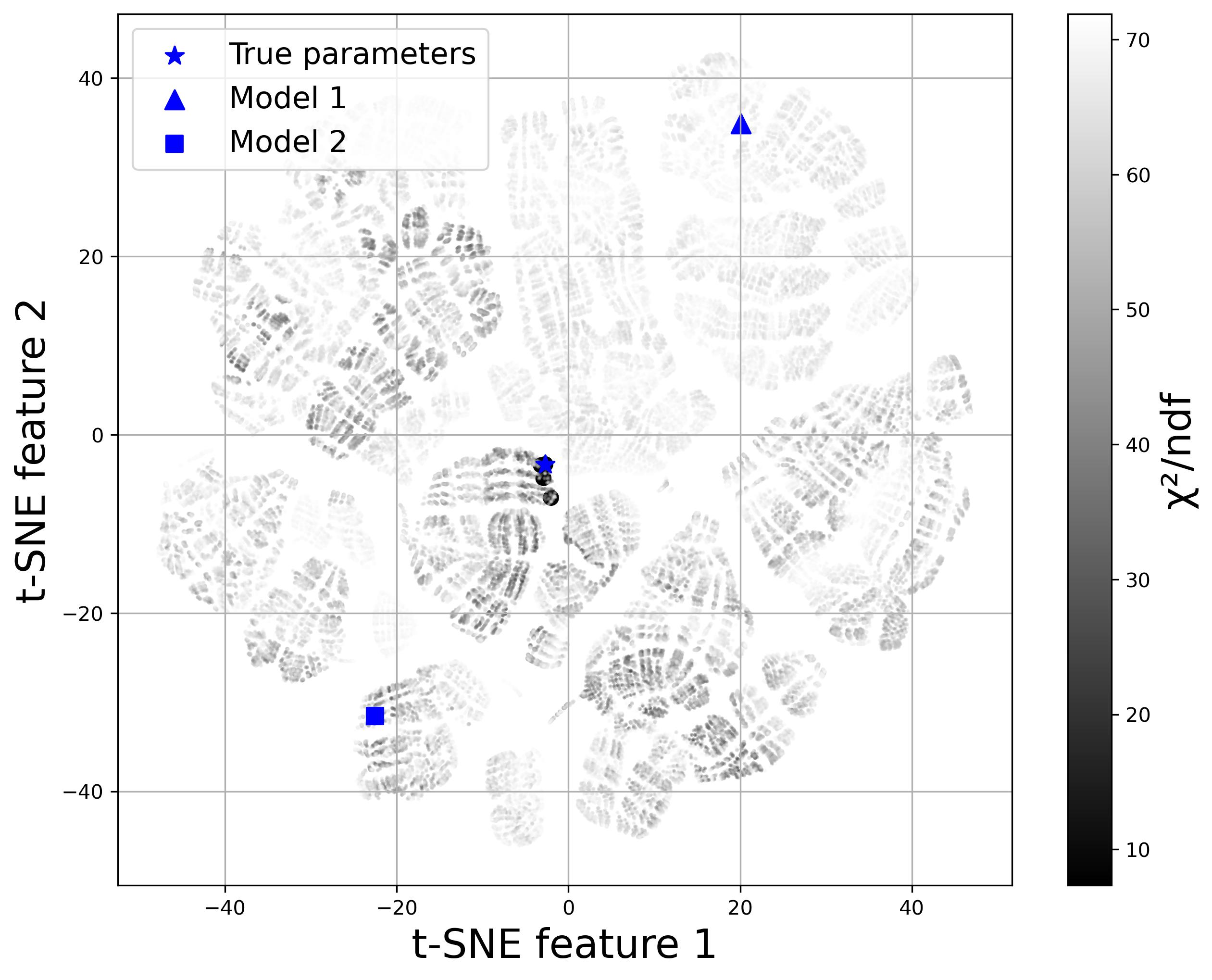}

     \end{subfigure}
      \begin{subfigure}{0.33\textwidth}
         \centering
         \includegraphics[width=\textwidth]{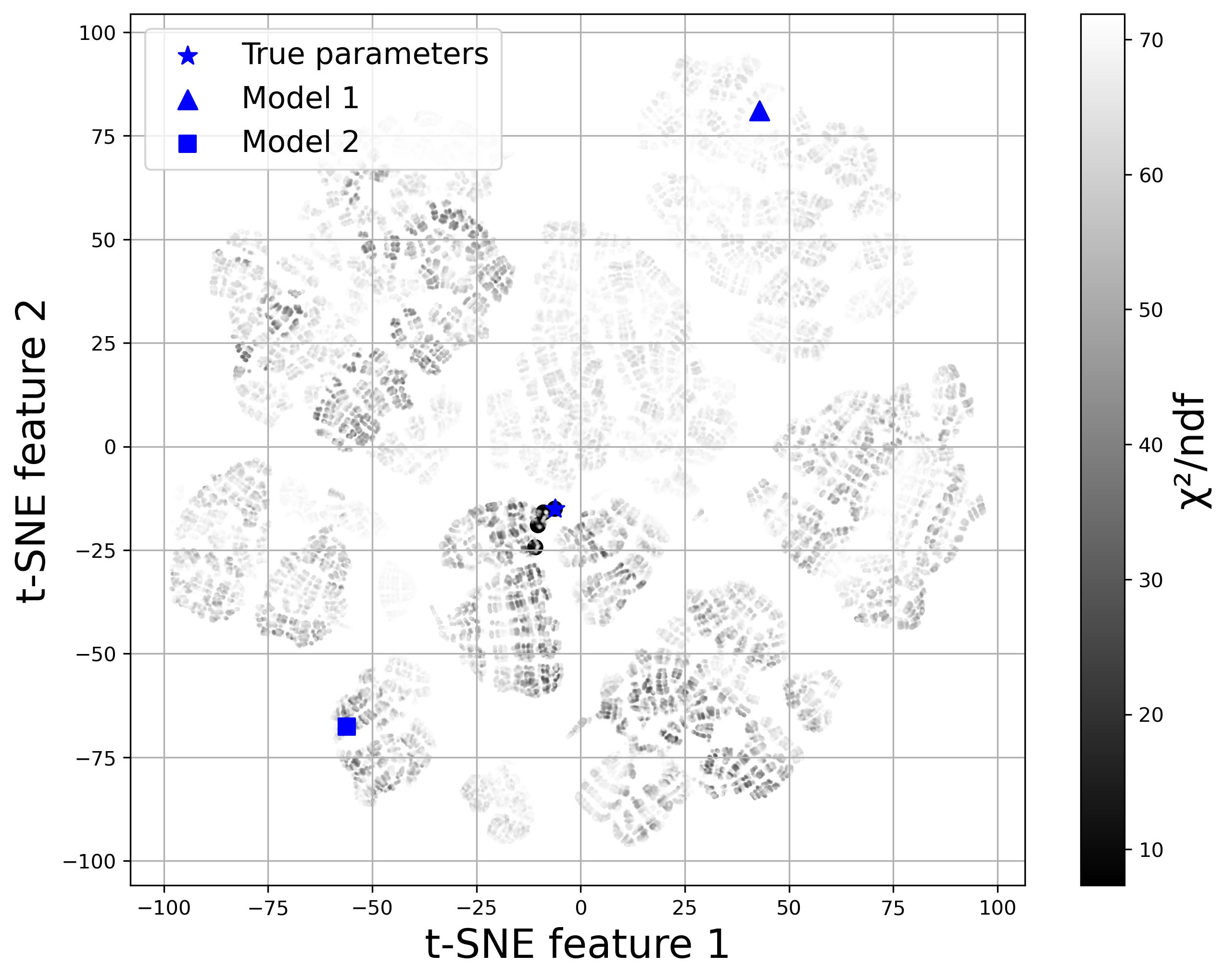}

     \end{subfigure}
        \caption{The plots show the impact of different learning rates. The perplexity is $p$ = 30 for all cases. The learning rate is $\eta$ = 10, 100 and 700 in this order.}
        \label{fig:learningrate}

\end{figure*}

We selected four different leptonic models (model parameters are shown in Table \ref{table:4sets}) and studied which hyperparameters lead to an adequate representation of the model parameter proximity in the parameter space. In the t-SNE plot, the first best solutions with a $\chi^{2}/$ndf < 9 are marked as black points. These solutions are similar to the true solution. One of them is the best solution resulting from the grid scan and is represented in the second row of Table \ref{table:4sets}. For this reason, t-SNE should locate the true solution in the same cluster as the marked points since we searched for a representation that would classify these points as similar. The third parameter set in the table has different parameters and should be placed in another cluster. The last dataset also has different parameters and belongs to the sets with the highest $\chi^{2}/$ndf. Fig. \ref{fig:perplexity} and Fig. \ref{fig:learningrate} show different cases of perplexity and learning rate. While the recommendation for the perplexity is between five and 50, the learning rate is supposed to be in the range [10.0, 1000.0]. We first tested different cases of perplexity with a fixed learning rate of 300. Figure \ref{fig:perplexity} shows the t-SNE plot for perplexity values $p$= 10, 30 and 80. The left plot shows the true solution in the same cluster as the best solutions and the different models are located in other clusters. But it is partially difficult to distinguish different clusters and everything seems to be merged together. The perplexity of 30 seems to be perfect since the different parameter sets are arranged as we intended. Clear clusters are visible which makes it possible to differentiate between different regions. The third plot shows the case of a perplexity $p$ = 80. The true solution is located in another cluster than the best solutions which gives the impression that these models could be very different. Therefore, we decided to select a perplexity of 30. If the learning rate is too low, the algorithm might get stuck in a local minimum instead of finding the global minimum of the Kullback-Leibler divergence. With a learning rate that is too high, the oscillations might be too high, and the algorithm could miss the global minimum. With a low learning rate of 10, we can notice a lack of structure and poorly separated points and clusters. In addition, the true solution is outside the cluster with the best solutions. The last two plots with a learning rate of 300 and 700 look almost similar. With a learning rate of 700, the true solution does not lie exactly on the best solutions and it indicates that they are similar but not identical. Due to that, we decided to select the learning rate of 700. Figure \ref{fig:learningrate} show three plots with different learning rates and a fixed perplexity of 30.

\begin{table*}
\caption{Parameters of true solution, best grid scan result, Model 1 and Model 2 in this order.}             
\label{table:4sets}      

\centering                          

\begin{tabular}{c c c c c c c}        

\hline\hline                 

lg($R_\mathrm{blob}^\prime$) [cm] & $B^\prime$ [G] & lg($\gamma_\mathrm{min}^\prime$) & lg($\gamma_\mathrm{max}^\prime$) & lg($L^\prime_\mathrm{e}~[erg~s^{-1}]$) & $\Gamma_\mathrm{b}$ & index $\alpha$  \\     
\hline                        

  16.58 & 0.41 & 3.30  & 4.88 & 43.39 & 7.79 & 2.82 \\      
   16.39  & 0.64 & 3.21    & 4.78 &43.11 & 9.00 & 2.83 \\
   16.67 & 3.37 & 3.63 & 4.00  & 44.22 & 3.00  & 1.83 \\
   16.11 & 0.10 & 3.00 & 4.67 & 42.56 & 27.00 & 3.50  \\
\hline

\end{tabular}
\end{table*}

\section{Boundaries of the parameter space}

\begin{table}[H]
    \centering
    \caption{Boundaries of the leptonic model parameters used in fitting simulated data.}
    \begin{tabular}{ll}
    \toprule
    Parameter & Value range\\
    \midrule
    $\log_{10} (R^\prime_\mathrm{blob})$, cm & [15.0, 17.5]\\
    $B^\prime$, gauss & [0.1, 5]\\
    $\Gamma_\mathrm{b}$ & [3.0, 30.0]\\
    $\log_{10} (\gamma_\mathrm{e}^{\prime\mathrm{min}})$ & [3.0, 4.0]\\
    $\log_{10} (\gamma_\mathrm{e}^{\prime\mathrm{max}})$ & [4.0, 5.0]\\
    $\alpha_\mathrm{e}$ & [0.5, 3.5]\\
    $\log_{10} (L^\prime_\mathrm{e})$,  erg s$^{-1}$ & [42.0, 47.0]\\
    \bottomrule
    \end{tabular}
    \tablefoot{Parameter description: R$'_{\mathrm{b}}$ is the radius of the blob, $B'$ is the magnetic field strength in the emission region, $\Gamma_{\mathrm{b}}$ is blob Lorentz factor; $\gamma'^{\textrm{min}}_{\mathrm{e}}$ and $\gamma'^{\textrm{max}}_{\mathrm{e}}$ are the minimum and maximum Lorentz factor of the electrons respectively, $\alpha_{\textrm{e}}$ is the spectral index of electrons, $L'_{\textrm{e}}$ is electron luminosity.}
    \label{tab:lep_pars_bound}
\end{table}

\begin{table}[H]
    \centering
    \caption{List of leptonic model parameters for Mrk 501.}
    \begin{tabular}{ll}
    \toprule
    Parameter & Value range\\
    \midrule
    $R^\prime_\mathrm{blob}$, cm & [$10^{15.5}$, $10^{18}$]\\
    $B^\prime$, gauss & [$10^{-3}$, 0.5]\\
    $\Gamma_\mathrm{b}$ & [5.0, 30.0]\\
    $\gamma_\mathrm{e}^{\prime\mathrm{min}}$ & [$10^{2}$, $10^{5}$]\\
    $\gamma_\mathrm{e}^{\prime\mathrm{max}}$ & [$10^{5}$, $10^{7}$]\\
    $\alpha_\mathrm{e}$ & [1.0, 3.0]\\
    $L^\prime_\mathrm{e}$ / erg s$^{-1}$ & [$10^{39}$,$10^{44}$]\\
    \bottomrule
    \end{tabular}
        \tablefoot{Parameter description: R$'_{\mathrm{b}}$ is the radius of the blob, $B'$ is the magnetic field strength in the emission region, $\Gamma_{\mathrm{b}}$ is blob Lorentz factor; $\gamma'^{\textrm{min}}_{\mathrm{e}}$ and $\gamma'^{\textrm{max}}_{\mathrm{e}}$ are the minimum and maximum Lorentz factor of the electrons respectively, $\alpha_{\textrm{e}}$ is the spectral index of electrons, $L'_{\textrm{e}}$ is electron luminosity.}
    
    \label{tab:lep_pars_grid_scan_mrk501}
\end{table}

\section{MCMC corner plots}
\label{sec:appendix:cornerplots}

\begin{figure}
    \centering

     \begin{subfigure}{0.45\textwidth}
         \centering
         \includegraphics[width=\linewidth]{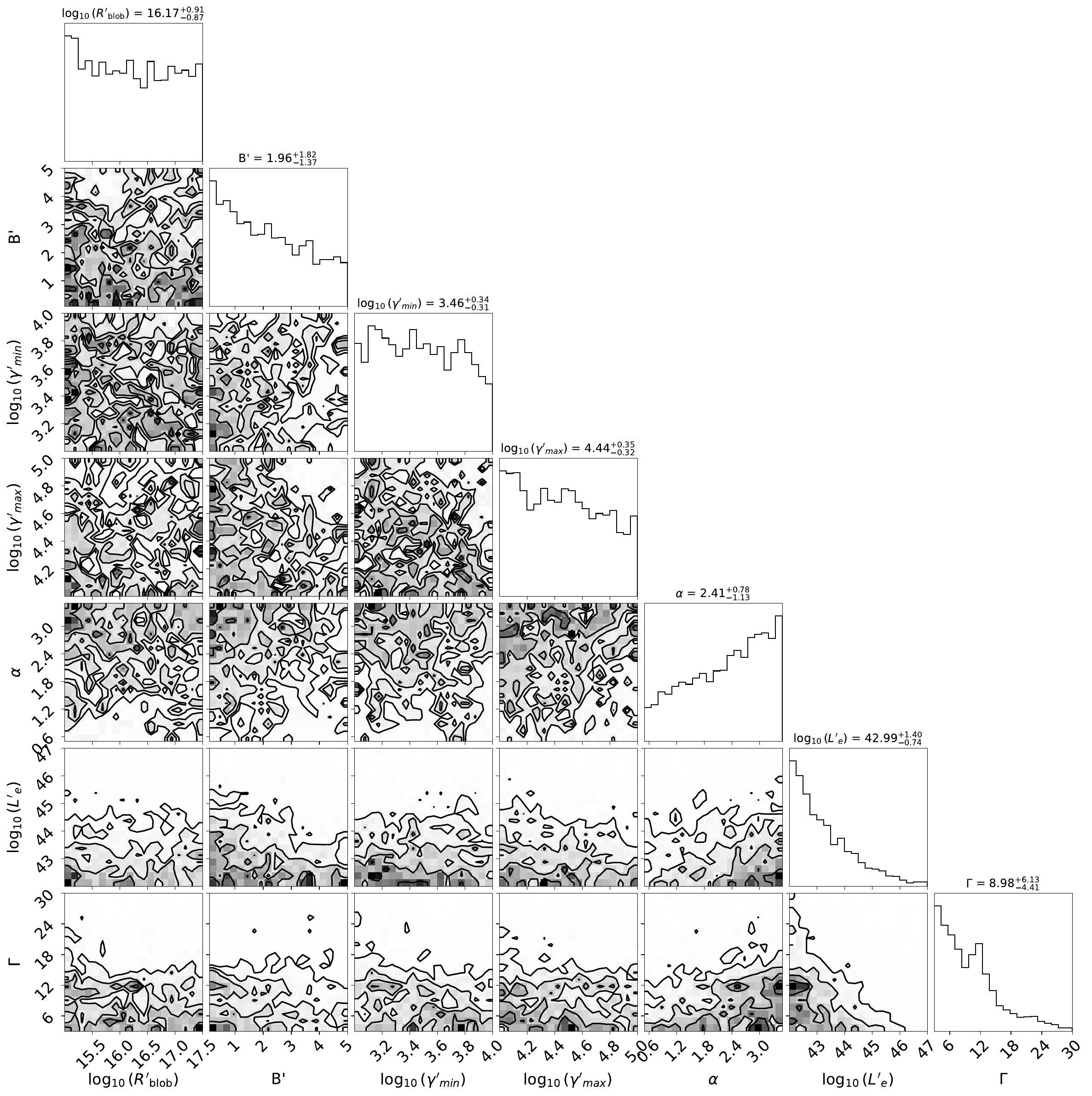}
         \caption{MCMC run 1.}
     \end{subfigure}

\begin{subfigure}{0.45\textwidth}
    \centering
    \includegraphics[width=\linewidth]{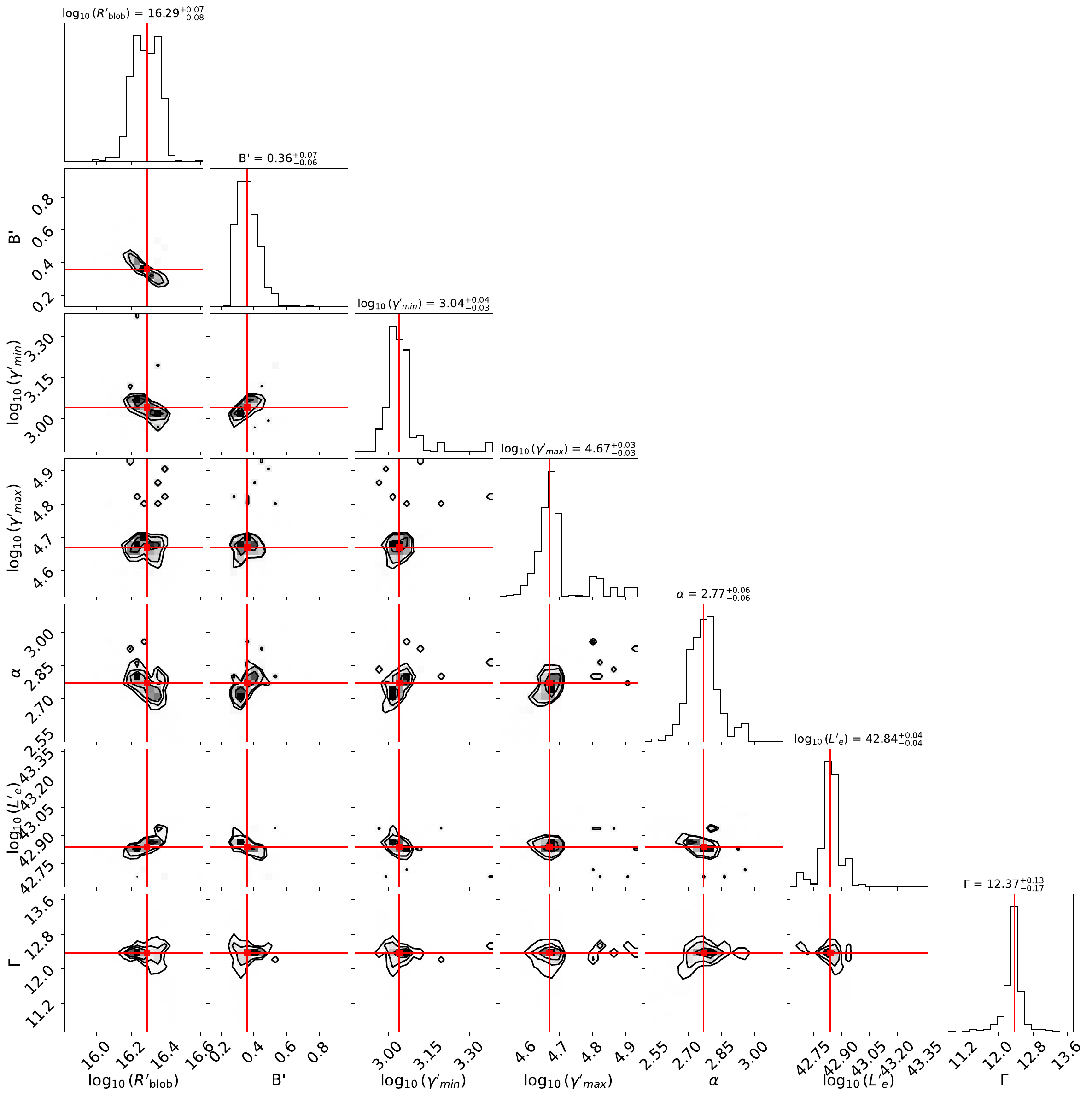}
    \caption{MCMC run 2.}
  
    \end{subfigure}
    
    \caption{Corner plot for simulated dataset 1.}
    \label{fig:MCMC_corner_ds1}
\end{figure}

\begin{figure}
    \centering

     \begin{subfigure}{0.45\textwidth}
         \centering
         \includegraphics[width=\linewidth]{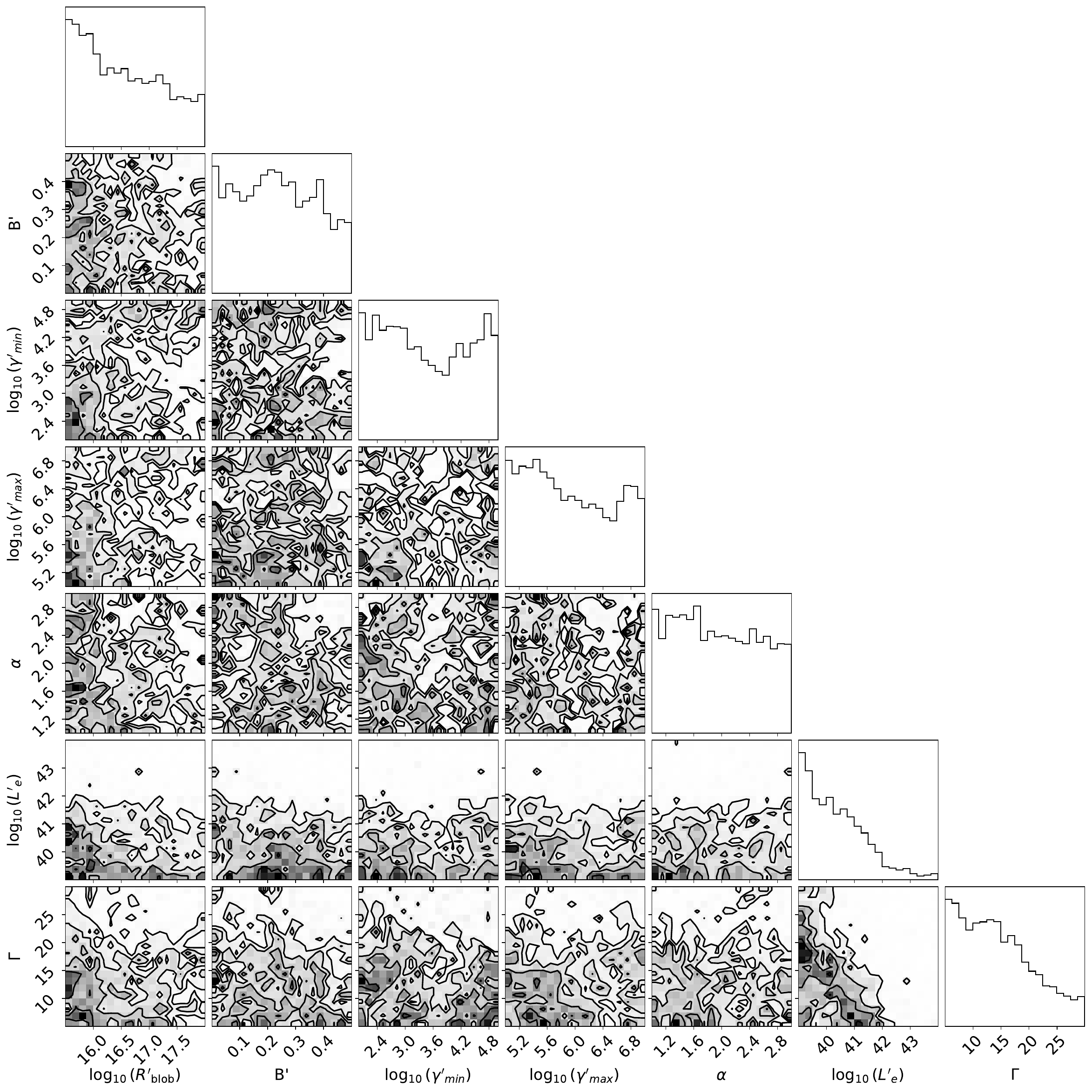}
         \caption{MCMC run 1.}
     \end{subfigure}

\begin{subfigure}{0.45\textwidth}
    \centering
    \includegraphics[width=\linewidth]{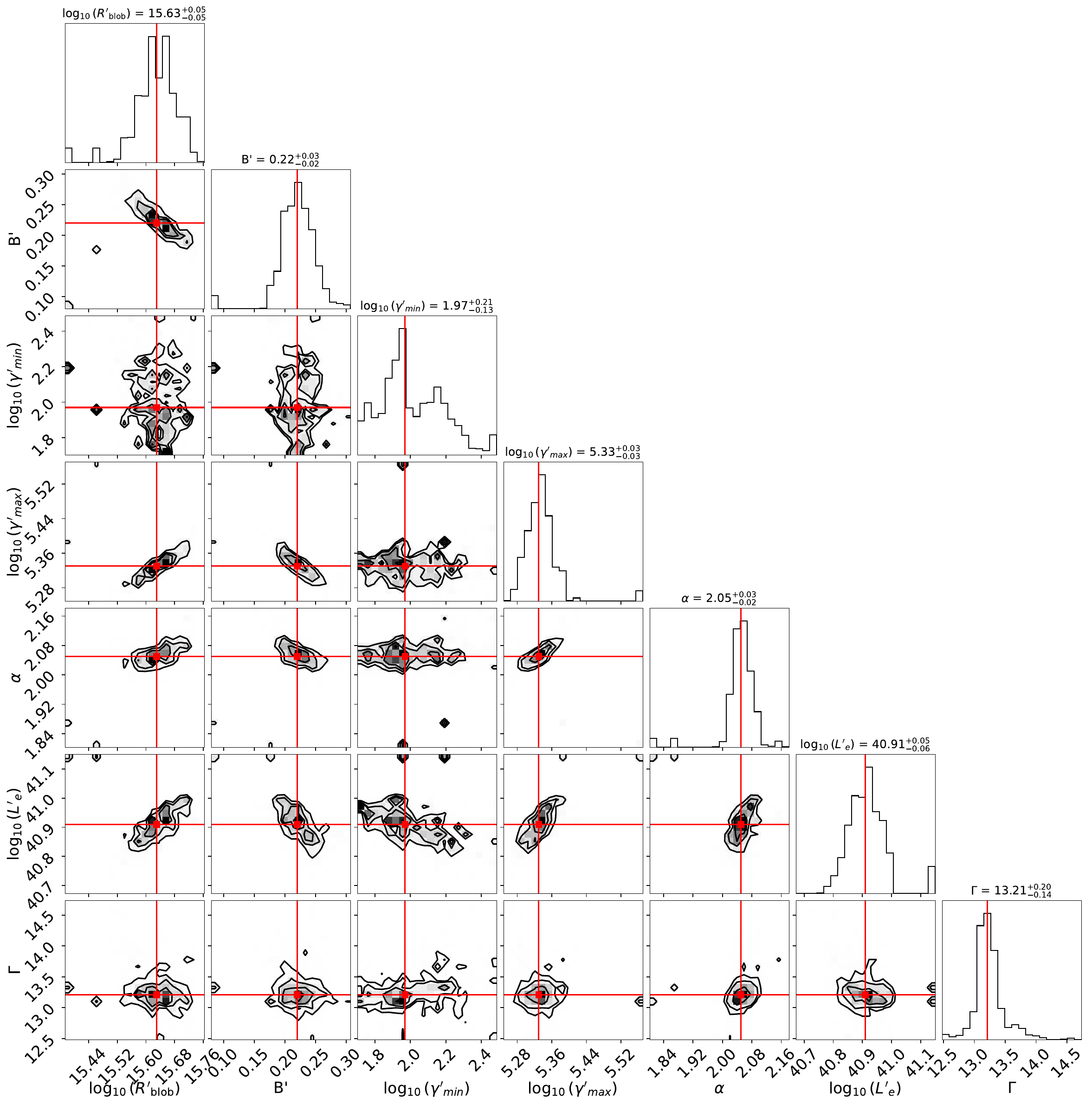}
    \caption{MCMC run 2.}
  
    \end{subfigure}
    
    \caption{Corner plot for Mrk 501. }
    \label{fig:mcmc_corner_mrk501}
\end{figure}

Figures \ref{fig:MCMC_corner_ds1} -- \ref{fig:mcmc_corner_mrk501} show the two-dimensional projections of the posterior probability distributions of the model parameters. The MCMC parameter search was completed in two steps, as described in Sec. \ref{sec:optimization}. The first step (run 1) corresponded to the global sampling within the boundaries defined for each SED. These boundaries were refined based on the results of the first step and the sampling was repeated (run 2). As shown in the plots for run 1 for all datasets, the selected number of walkers and steps was not enough for the algorithm to converge to one solution in the global parameter space. However, in the limited region, the convergence was achieved and the solution was found as demonstrated with the plots for run 2.

\begin{figure}
    \centering

     \begin{subfigure}{0.45\textwidth}
         \centering
         \includegraphics[width=\linewidth]{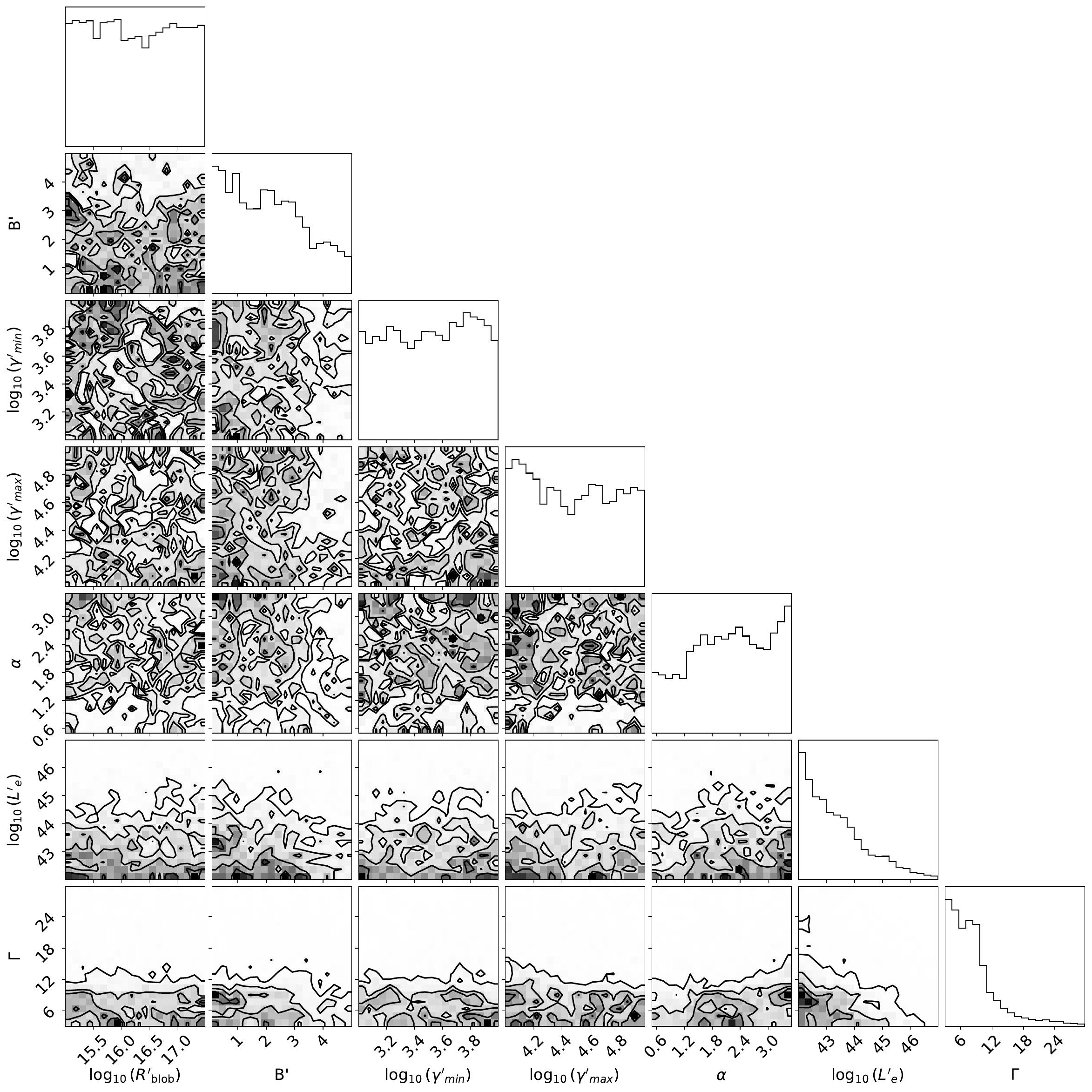}
         \caption{MCMC run 1.}
     \end{subfigure}

\begin{subfigure}{0.45\textwidth}
    \centering
    \includegraphics[width=\linewidth]{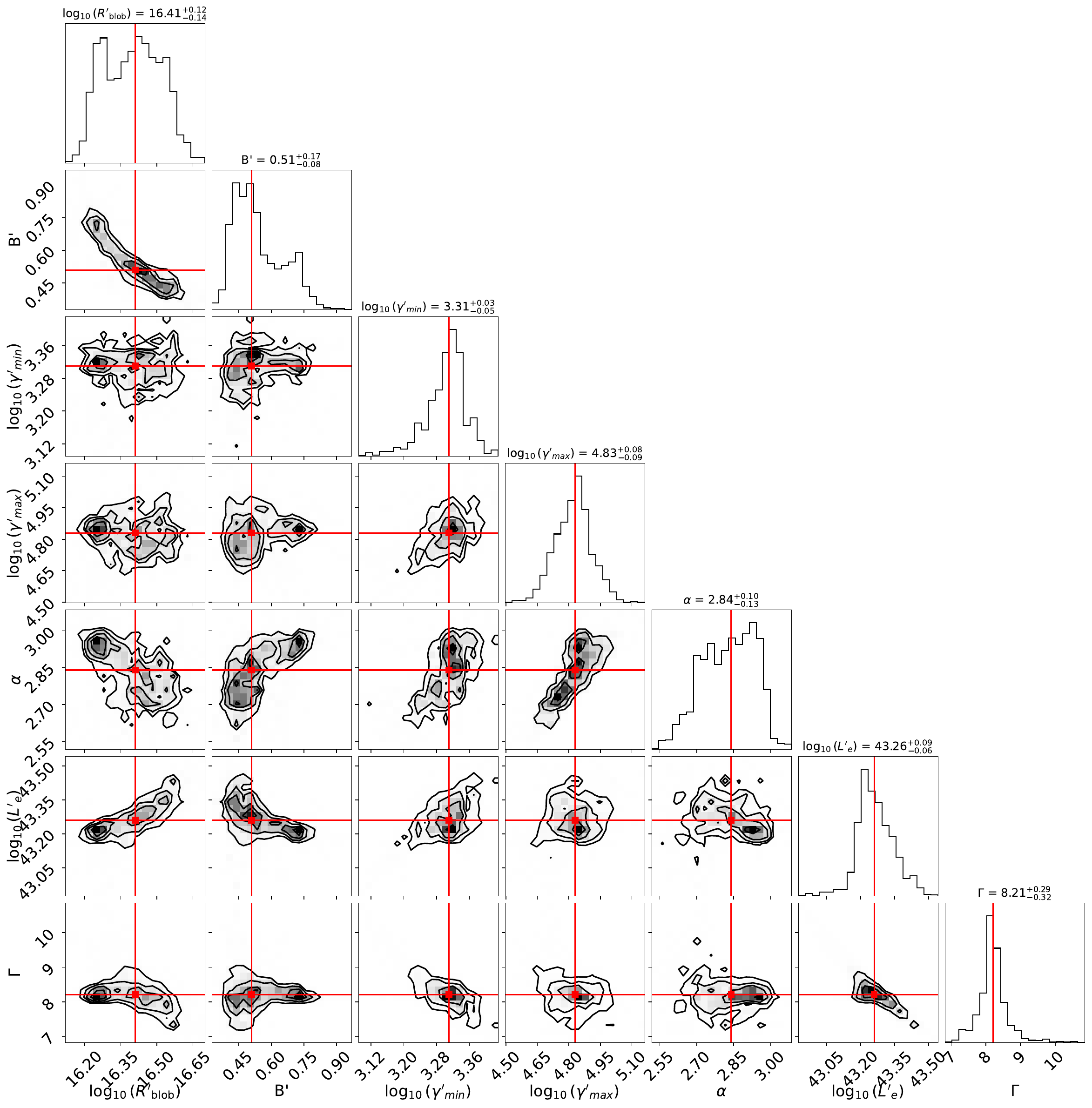}
    \caption{MCMC run 2.}
    \end{subfigure}
    \caption{Corner plot for simulated dataset 2.}
\end{figure}

\begin{figure}
    \centering

     \begin{subfigure}{0.45\textwidth}
         \centering
         \includegraphics[width=\linewidth]{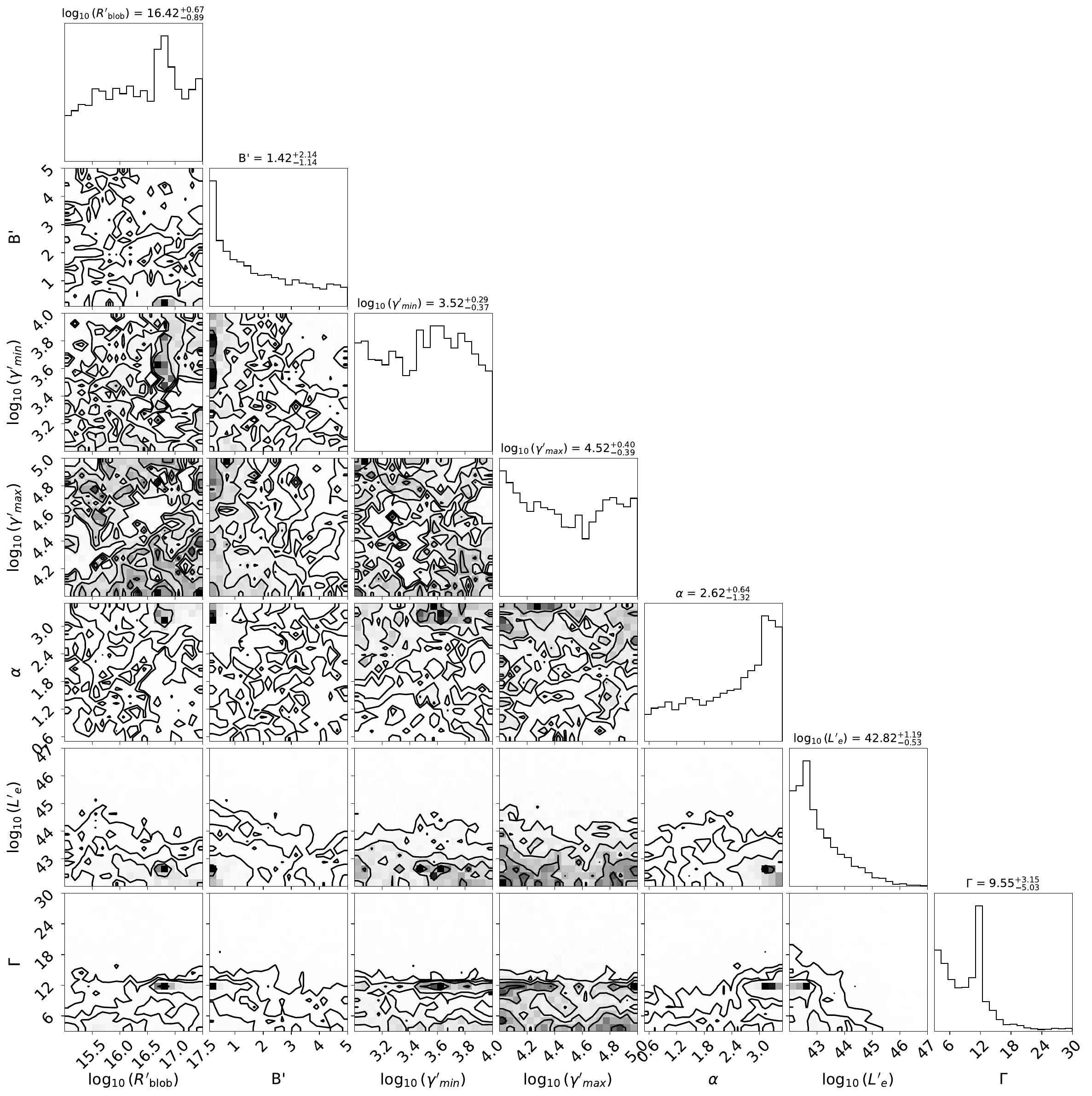}
         \caption{MCMC run 1.}
     \end{subfigure}

\begin{subfigure}{0.45\textwidth}
    \centering
    \includegraphics[width=\linewidth]{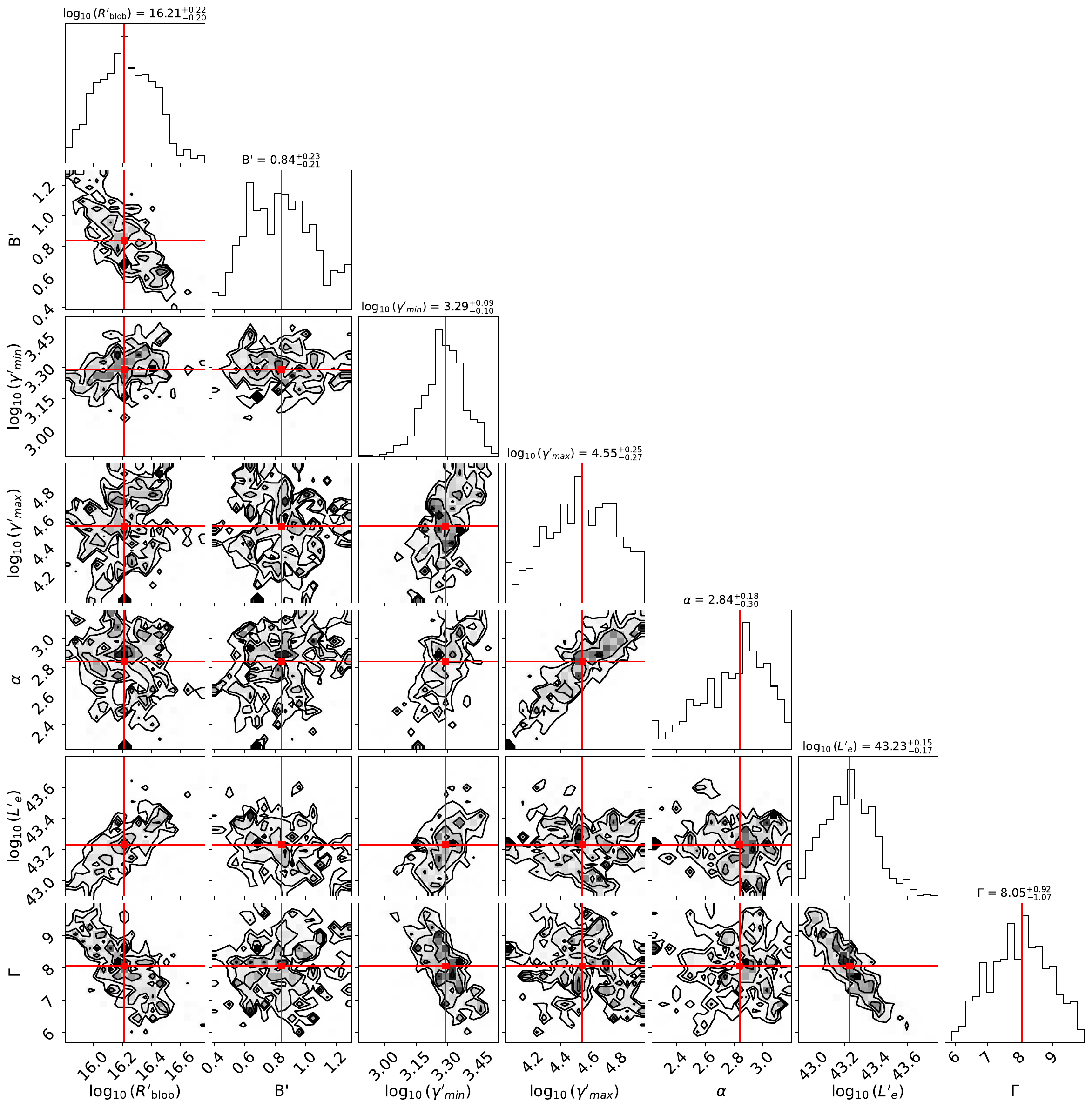}
    \caption{MCMC run 2.}
  
    \end{subfigure}
    \caption{Corner plot for simulated dataset 3.}
\end{figure}

\begin{figure}
    \centering

     \begin{subfigure}{0.45\textwidth}
         \centering
         \includegraphics[width=\linewidth]{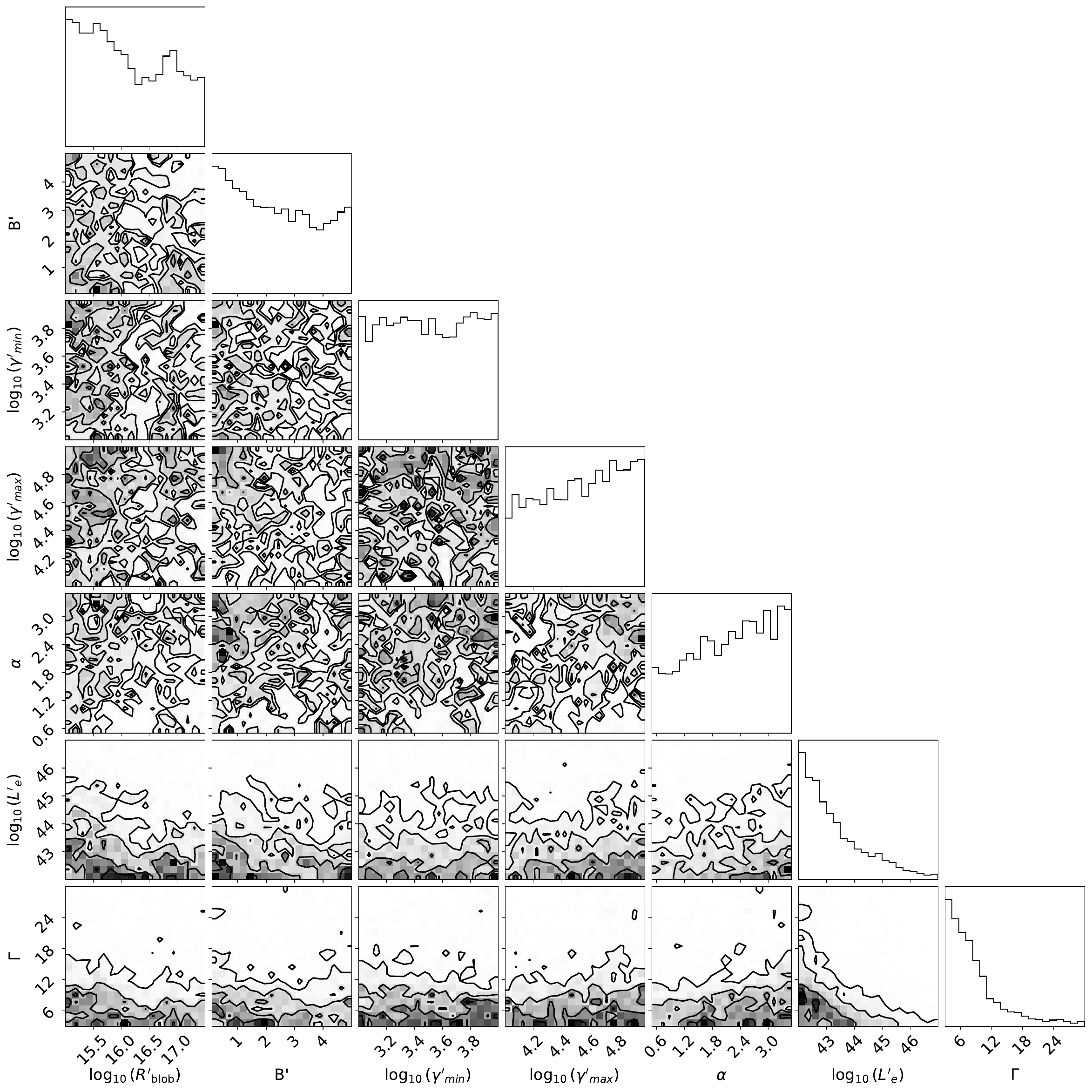}
         \caption{MCMC run 1.}
     \end{subfigure}

\begin{subfigure}{0.45\textwidth}
    \centering
    \includegraphics[width=\linewidth]{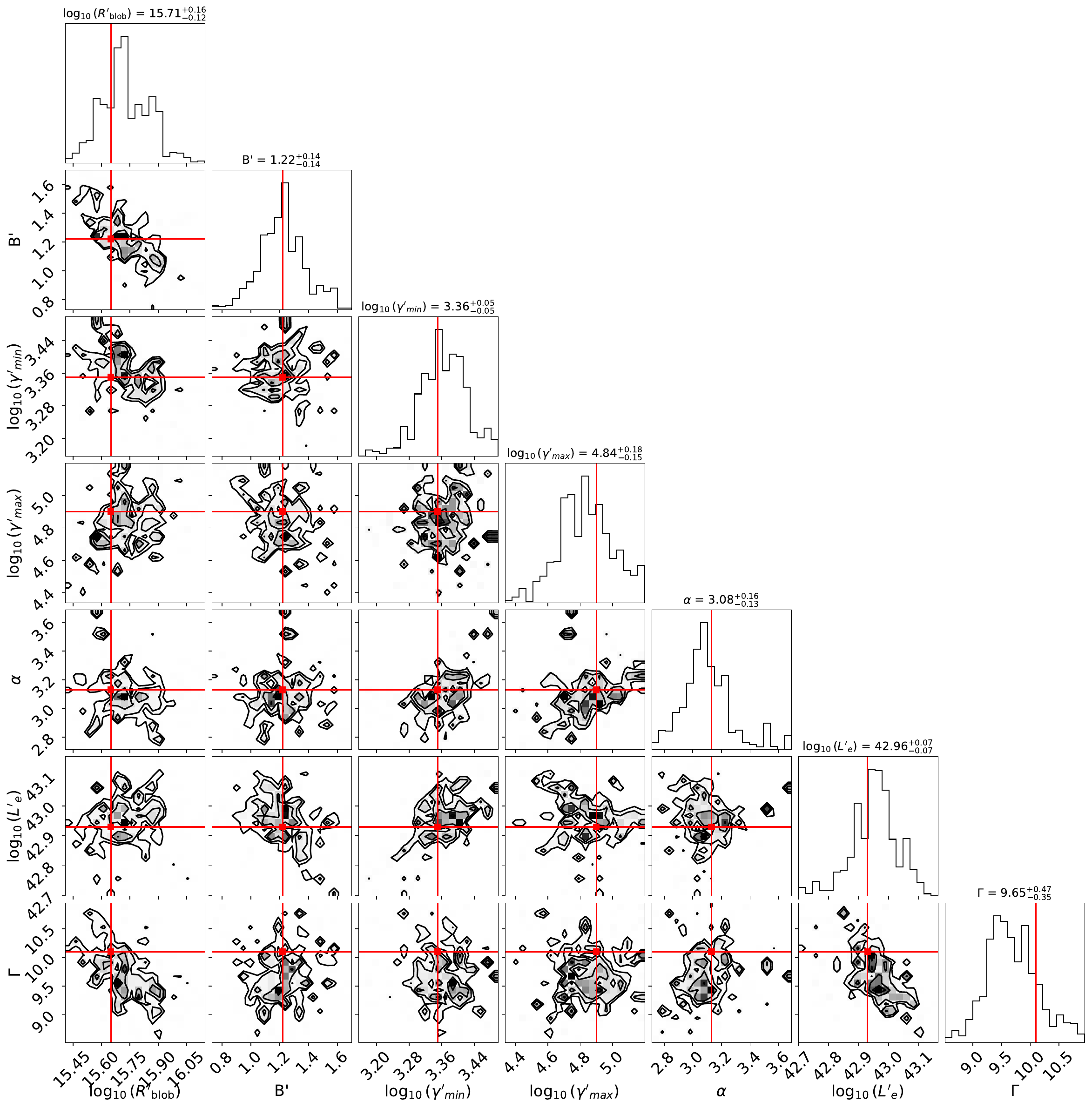}
    \caption{MCMC run 2.}
  
    \end{subfigure}
    \caption{Corner plots for PKS 0735+178.}
\end{figure}

\section{Comparison of optimization algorithms.}
\begin{table*}
    \centering
    \caption{Overview and comparison of different algorithms}

    \begin{tabular}{p{0.15\linewidth}p{0.07\linewidth}p{0.33\linewidth}p{0.35\linewidth}}
    \toprule
    Algorithm & N$_{\textrm{models}}$ & Advantage & Disadvantage \\
    \midrule
    Grid scan with local minimization & $10^{7}$ & Considers more than only one solution, can be reused for other sources, few controllable parameters & Computationally expensive, drastic increase in computing cost with the increase of model parameters\\
    
    Genetic algorithm &  $10^{5} - 10^{6}$& Able to cover large part of parameter space, overcomes local minima & Only one solution, sensitive to setting of algorithm parameters \\
    CMA-ES & $10^{5}$ & Able to cover large part of parameter space, overcomes local minima, self-adaption of algorithm parameters & Only one solution \\
    Minuit (\texttt{simplex} + \texttt{migrad}) & $10^{3}$ & Computationally inexpensive, few controllable parameters & May be sensitive to local minima, only one solution, sensitive to the choice of initial point \\
    MCMC (ensemble sampler) & $10^{5}$ & Provides parameter distributions & Sensitive to the choice of initial point and algorithm parameters \\

    \bottomrule
    \end{tabular}
    \tablefoot{N$_{\textrm{models}}$ is the total number of generated models in the selected algorithm.}
    \label{tab:performance}
\end{table*}

\begin{figure} [htbp!]
    \centering
    \includegraphics[width=0.9\linewidth]{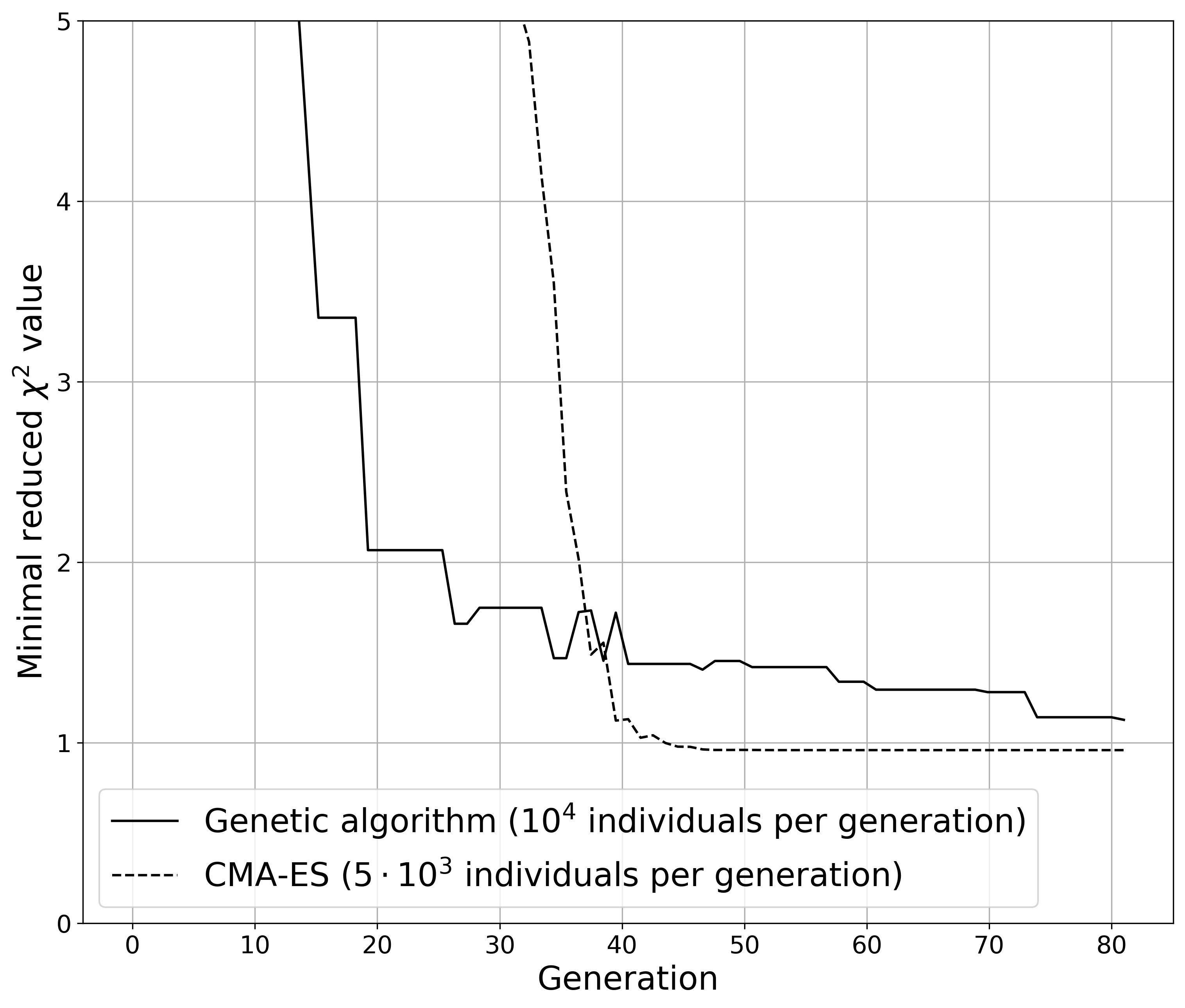}
    \caption{Comparison of the convergence between the genetic algorithm and the CMA-ES. The plot shows the smallest reduced $\chi^{2}$ of each generation depending on the number of the generation.}
    \label{fig:convergence}
\end{figure}
\end{appendix}
\end{document}